\documentclass[aps,prx,notitlepage,superscriptaddress,longbibliography,twocolumn,nofootinbib,floatfix]{revtex4-2} 

\usepackage[utf8]{inputenc}
\usepackage{graphicx}
\usepackage{hyperref}
\usepackage[dvipsnames]{xcolor}
\usepackage{graphicx}

\usepackage{tikz,tikz-3dplot}
\usetikzlibrary{decorations.pathreplacing}
\usepackage{pgfplots}

\usepackage{lipsum}

\usepackage{amsmath,amsthm,amssymb,mathtools}
\usepackage{multirow}
\usepackage{braket}
\usepackage{bbm,bm}
\usepackage[bb=boondox]{mathalfa}

\DeclareFontFamily{U}{matha}{\hyphenchar\font45}
\DeclareFontShape{U}{matha}{m}{n}{
      <5> <6> <7> <8> <9> <10> gen * matha
      <10.95> matha10 <12> <14.4> <17.28> <20.74> <24.88> matha12
      }{}
\DeclareSymbolFont{matha}{U}{matha}{m}{n}
\DeclareMathSymbol{\muparrow}{3}{matha}{"D2}
\DeclareMathSymbol{\mdownarrow}{3}{matha}{"D3}
\DeclareMathSymbol{\mupdownarrow}{3}{matha}{"D9}

\DeclareFontFamily{U}{mathb}{\hyphenchar\font45}
\DeclareFontShape{U}{mathb}{m}{n}{
      <5> <6> <7> <8> <9> <10> gen * mathb
      <10.95> mathb10 <12> <14.4> <17.28> <20.74> <24.88> mathb12
      }{}
\DeclareSymbolFont{mathb}{U}{mathb}{m}{n}
\DeclareMathSymbol{\mdownuparrows}{3}{mathb}{"D7}

\usepackage{makecell}

\usepackage{float}

\definecolor{dgreen}{rgb}{0,0.666,0}
\definecolor{dorange}{rgb}{0.666,.333,0}

\newcommand{\ie}{{ i.e.},\ }
\newcommand{\eg}{{ e.g.},\ }

\newcommand{\Tr}{\operatorname{Tr}}

\makeatletter
\pgfmathdeclarefunction{erf}{1}{%
  \begingroup
    \pgfmathparse{#1 > 0 ? 1 : -1}%
    \edef\sign{\pgfmathresult}%
    \pgfmathparse{abs(#1)}%
    \edef\x{\pgfmathresult}%
    \pgfmathparse{1/(1+0.3275911*\x)}%
    \edef\t{\pgfmathresult}%
    \pgfmathparse{%
      1 - (((((1.061405429*\t -1.453152027)*\t) + 1.421413741)*\t 
      -0.284496736)*\t + 0.254829592)*\t*exp(-(\x*\x))}%
    \edef\y{\pgfmathresult}%
    \pgfmathparse{(\sign)*\y}%
    \pgfmath@smuggleone\pgfmathresult%
  \endgroup
}
\makeatother

\begin{document}
\title{Probing Hilbert space fragmentation and the block inverse participation ratio}

\author{Philipp Frey}
\affiliation{School of Physics, The University of Melbourne, Parkville, VIC 3010, Australia}

\author{David Mikhail}
\affiliation{School of Physics, The University of Melbourne, Parkville, VIC 3010, Australia}

\author{Stephan Rachel}
\affiliation{School of Physics, The University of Melbourne, Parkville, VIC 3010, Australia}

\author{Lucas Hackl}
\affiliation{School of Mathematics and Statistics, The University of Melbourne, Parkville, VIC 3010, Australia}
\affiliation{School of Physics, The University of Melbourne, Parkville, VIC 3010, Australia}

\begin{abstract}
We consider a family of quantum many-body Hamiltonians that show exact Hilbert space fragmentation in certain limits. The question arises whether fragmentation has implications for Hamiltonians in the vicinity of the subset defined by these exactly fragmented models, in particular in the thermodynamic limit. We attempt to illuminate this issue by considering distinguishable classes of transitional behavior between fragmented and nonfragmented regimes and employing a set of numerical observables that indicate this transition. As one of these observables we present a modified inverse participation ratio (IPR) that is designed to capture the emergence of fragmented block structures. We compare this block IPR to other definitions of inverse participation ratios, as well as to the more traditional measures of level-spacing statistics and entanglement entropy. In order to resolve subtleties that arise in the numerics, we use perturbation theory around the fragmented limit as a basis for defining an effective block structure. We find that our block IPR predicts a boundary between fragmented and nonfragmented regimes that is compatible with results based on level statistics and bipartite entanglement. A scaling analysis indicates that a finite region around the exactly fragmented limit is dominated by effects of approximate fragmentation, even in the thermodynamic limit, and suggests that fragmentation constitutes a phase. We provide evidence for the universality of our approach by applying it to a different family of Hamiltonians, that features a fragmented limit due to emergent dipole conservation.
\end{abstract}

\maketitle


\clearpage


\section{Introduction}

Dynamical thermalization and ergodicity breaking in closed or driven many-body quantum systems have been subjects of renewed interest in recent years \cite{de2019dynamics,Moudgalya_2022,doi:10.1146/annurev-conmatphys-031214-014726,PhysRevB.92.134204,PhysRevB.82.174411,Smith2016,PhysRevLett.95.206603,PhysRevB.76.052203,PhysRevB.75.155111,PhysRevB.98.174202,PhysRevB.88.014206,Schreiber842,Choi1547,PhysRevLett.121.023601,PhysRevLett.117.040601,PhysRevX.9.021003,PhysRevX.10.011047,PhysRevB.101.174204,Wilczek_2012,RevModPhys.91.021001,PhysRevLett.118.030401,PhysRevB.94.085112,srednicki-94pre888,khemani-16prl250401,kvorning_time-evolution_2022,doi:10.1126/sciadv.abm7652}, with a particular focus on possible exceptions to what seems to be the generic mechanism of thermalization. A thermal Hamiltonian is characterized by eigenstates that look thermal with respect to local observables and have additional properties that ensure that arbitrary initial states thermalize, as long as they are not unphysical. These properties are summarized in what is called the {\it eigenstate thermalization hypothesis} (ETH) \cite{srednicki-94pre888,deutsch91pra2046,rigol-08n854}. One of the proposed exceptions to ETH is {\it many-body localization} (MBL) \cite{ALET2018498,PhysRevLett.116.247204,PhysRevLett.118.017201,PhysRevX.7.021018,doi:10.1146/annurev-conmatphys-031214-014701,PhysRevLett.113.107204,PhysRevB.91.081103,PhysRevB.75.155111, PhysRevB.82.174411, RevModPhys.91.021001, ALET2018498, PhysRevB.91.081103, PhysRevLett.113.107204, doi:10.1146/annurev-conmatphys-031214-014701,PhysRevB.98.174202}, which can be understood as an emergent integrability caused by local quenched disorder. The question of whether MBL constitutes a phase in the sense of robustness against a certain class of or even arbitrary perturbations remains somewhat controversial \cite{ABANIN2021168415,leblond2021universality,PhysRevB.105.174205,vznidarivc2018interaction,Doggen_2021,sierant2020thouless,vsuntajs2020ergodicity,panda2020can,PhysRevB.98.024203,PhysRevLett.124.243601,PhysRevB.103.024203}. While there is a proposed proof of robustness in one dimension under certain extra assumptions \cite{Imbrie2016}, its validity has recently been called into question by a number of numerical results \cite{PhysRevE.102.062144,PhysRevE.104.054105}. In dimensions $d>1$ MBL is generally considered to be unstable but there are proposed exceptions \cite{PhysRevResearch.5.L032011}.

In order to identify a many-body localized Hamiltonian and distinguish it from a thermal one, several different numerical probes may be used. One of these probes is the level statistics of the Hamiltonian's spectrum. The spacing of eigenvalues within a thermal regime is well described by the {\it Gaussian orthogonal ensemble} (GOE) from random matrix theory, meaning the set of random Hermitian matrices with Gaussian distributed entries \cite{PhysRevLett.52.1,de_marco_level_2022}. Within an integrable or MBL regime the level-spacing matches the Poisson statistics of independent random levels \cite{PhysRevB.93.041424,PhysRevE.61.6278}. Another measure is the entanglement entropy of eigenstates with respect to real-space partitions of the system. Eigenstates that satisfy ETH are predicted to have volume-law entanglement \cite{PhysRevX.8.021026,bianchi2021volume,PhysRevE.97.012140,HUANG2019594}, while the MBL case should yield area-law entanglement \cite{PhysRevLett.111.127201,Bauer_2013,RevModPhys.91.021001}. A third quantity that is often used in the context of MBL is the \textit{inverse participation ratio} (IPR) \cite{murphy_generalized_2011,PhysRevA.93.042101}, which is a basis-dependent measure of the randomness of eigenstates.

In this paper we consider Hilbert space fragmentation \cite{PhysRevB.101.214205,PhysRevX.12.011050,Moudgalya_2021,PhysRevX.10.011047,PhysRevB.101.174204,PhysRevB.103.L220304,Moudgalya_2022, PhysRevLett.127.150601, PhysRevX.9.021003, PhysRevB.103.L220304,Dias_2000,de2019dynamics,frey_hilbert_2022,PhysRevX.10.011047,yoshinaga_emergence_2022,PhysRevLett.130.010201}, an entirely different mechanism of ergodicity breaking, which is much less explored than MBL. It describes the phenomenon that local constraints can separate the Hilbert space into exponentially many subspaces, which are spanned by product states and are dynamically disconnected from each other. The Hamiltonian assumes a block-diagonal structure with respect to a product basis, and the number of blocks is exponential in system size~$L$. This results in a small number of dynamically accessible states and can prevent a typical initial product state (or their low-dimensional linear combinations) from thermalizing. As with the case of MBL, it is not entirely clear how robust this mechanism of ergodicity breaking is. While some argue heuristically that for physical realizations the fragmentation is merely a prethermal phenomenon even in the thermodynamic limit \cite{PhysRevB.101.174204}, others have demonstrated the possibility of much greater robustness \cite{PhysRevB.108.045127,stephen2022ergodicity}. Instead of arguing in terms of the lifetime of dynamical constraints based on perturbative effects, we choose to study properties of the exact eigenstates of models that have a fragmented limit as well as nonfragmented regimes. In particular, we introduce a modified notion of IPR in order to quantify the degree of fragmentation indicated by these eigenstates.


\subsection{Stability of fragmentation with respect to finite perturbations}

If we consider the parameter space describing locally interacting Hamiltonians, those Hamiltonians exhibiting Hilbert space fragmentation form a low-dimensional subset. We can draw an analogy to non-diagonalizable matrices, which form a similar low-dimensional subset within the space of all square matrices of a given dimension. Technically, as soon as we leave the respective subset, embedded as a hypersurface in the larger space, the Hilbert space ceases to be fragmented, just in the same way as almost all matrices become diagonalizable.

\begin{figure}[t]
    \centering
    \begin{tikzpicture}

    \begin{scope}[yshift=-.3cm,xshift=.3cm]
    \fill[blue,opacity=.2] (0,-3.5) rectangle (3.73,-4.1);
    \draw[black,dashed,ultra thick] (3.75,-3.5) -- (3.75,-4.1) node[below]{$J^*$};
    \draw[ultra thick,Blue] (0,-3.5) node[xshift=-3mm,yshift=.3cm,right,black]{\textbf{Example 1: $\hat{H}_{\mathrm{dis}}+J\,\hat{H}_{\mathrm{tb}}$ (MBL)}} -- (0,-4.1) node[below]{$0$};
    \draw[red,ultra thick,->] (0,-3.8) -- (7.5,-3.8) node[below]{$J$};
    \draw (1.875,-4.1) node[below]{MBL} (5.625,-4.1) node[below]{Thermal};
    \end{scope}

    \begin{scope}[yshift=-2.3cm,xshift=.3cm]
    \fill[blue,opacity=.2] (0,-3.5) rectangle (3.73,-4.1);
    \draw[black,dashed,ultra thick] (3.75,-3.5) -- (3.75,-4.1) node[below]{$\epsilon^*$};
    \draw[ultra thick,Blue] (0,-3.5) node[xshift=-3mm,yshift=.3cm,right,black]{\textbf{Example 2: $\hat{H}_{\mathrm{frag}}+\epsilon \hat{H}_{\mathrm{pert}}$ (Fragmented)}} -- (0,-4.1) node[below]{$0$};
    \draw[red,ultra thick,->] (0,-3.8) -- (7.5,-3.8) node[below]{$\epsilon$};
    \draw (1.875,-4.1) node[below]{Fragmented} (5.625,-4.1) node[below]{Thermal};
    \end{scope}
        
    \draw (0,0) rectangle (8,-3);
    \draw (-.05,.2) node[right]{Parameter space of Hamiltonians with local interaction};

    \clip(0,0) rectangle (8,-3);
    \begin{scope}[xshift=-1cm,yshift=-3cm]
    \begin{axis}[
    hide axis,
    domain=0:120,
    samples=120,width=12cm,height=4.1cm]
    \addplot[black,dashed,domain=0:120,ultra  thick, double=white,double   distance=10pt] {.5*sin(x*5.34/1.3)+.5*cos(x/1.3)};
    \addplot[white,domain=0:120,line width=10pt] {.5*sin(x*5.34/1.3)+.5*cos(x/1.3)};
    \addplot[blue,opacity=0.2,domain=0:120,line width=10pt] {.5*sin(x*5.34/1.3)+.5*cos(x/1.3)};
    \addplot[Blue,domain=0:120,ultra  thick, no marks] {.5*sin(x*5.34/1.3)+.5*cos(x/1.3)};
    \addplot[red,domain=41.4:48,ultra  thick, no marks,->] {-.5*sin(x*5.34/1.3)-.5*cos(x/1.3)+1};
    \end{axis} 
    \end{scope}

    \draw[Blue,thick] (2.2,-1) -- (2.1,-.4) node[right]{Hamiltonians satisfying exact condition};
    \draw[red] (3,-.9) node[right]{1-dimensional probe};
    \draw[] (1,-1) -- (.4,-2.5) node[right]{Effective region};
       
    \end{tikzpicture}
    \caption{We illustrate the basic idea of studying classes of Hamiltonians and their properties near subsets with special properties (many-body localization, Hilbert space fragmentation). Here, $\hat{H}_{\mathrm{dis}}+J\,\hat{H}_{\mathrm{tb}}$ represents the Hamiltonian (\ref{eq:MBL-Hamiltonian}).}
    \label{fig:space_of_hamiltonians}
\end{figure}
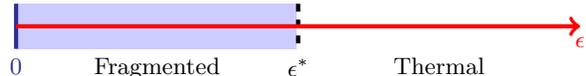

In practice, we expect that there is a separation of scales and as long as one is close enough to the respective subset one expects to still observe the physical phenomenon of fragmentation, just in the same way as a nearly non-diagonalizable matrix will appear to be ill conditioned for numerical diagonalization routines. It is therefore a natural question how we can probe Hilbert space fragmentation perturbatively in such a way that we can capture a region around the strictly fragmented hypersurface, which still shows the physical phenomena of fragmented evolution, even if the respective Hamiltonians do not give rise to exact Hilbert space fragmentation in the mathematical sense. Intuitively, it should be clear that even if time evolution technically connects all product states within a symmetry sector, the dynamics will appear to be fragmented if a time-evolved initial state is overwhelmingly restricted to a quasifragmented subspace with only little overlap with the rest of the Hilbert space.

To make this intuitive picture precise and quantifiable, we study three probes of fragmentation---an appropriately defined inverse participation ratio, bipartite entanglement entropy, and level-spacing statistics---which allow us to identify an effective region of quasi-Hilbert space fragmentation around some well-known fragmented models. For concrete calculations, we focus on one- and two-dimensional subspaces in the parameter space as illustrated in Fig.~\ref{fig:space_of_hamiltonians}. Consequently, our paper is largely based on numerical studies of specific models, which we consider to be representative for most known examples of fragmentation. We also develop the notion of an effective block structure and explain how to construct it in general for any fragmented model with exact symmetries such as translation invariance.

\begin{figure}
    \centering
    \begin{tikzpicture}
    
    \begin{scope}[yshift=2.3cm]
    \draw (2,-5) node{$f(\epsilon)$};
    \draw (2.8,-5.7) node[red]{$\epsilon$};
    \draw[red,ultra thick,->] (0,-6) -- (3,-6);
    \draw[black, very thick,->] (0,-6) -- (0,-4.5);
    \draw[gray,thick] plot[domain=0:3,samples=200] (\x,{erf(-(\x-1.5)*3)/2-5.5});
    \draw[dashed,blue,thick] plot[domain=0:3,samples=200] (\x,{erf(-(\x-1.5)*3)/2-5.5});
    \end{scope}

    \begin{scope}[yshift=.5cm,xshift=4.5cm,yscale=.5]
    \draw[red,ultra thick,->] (0,-6) -- (3,-6);
    \draw[black, very thick,->] (0,-6) -- (0,-4.5);
    \draw[gray,thick] plot[domain=0:3,samples=200] (\x,{erf(-(\x-1.5)*50)/2-5.5});
    \draw[dashed,blue,thick] plot[domain=0:3,samples=200] (\x,{erf(-(\x-1.5)*50)/2-5.5});
    \end{scope}
    \begin{scope}[yshift=-.5cm,xshift=4.5cm,yscale=.5]
    \draw[red,ultra thick,->] (0,-6) -- (3,-6);
    \draw[black, very thick,->] (0,-6) -- (0,-4.5);
    \draw[gray,thick] plot[domain=0:3,samples=200] (\x,{erf(-(\x-1.5)*5)/2-5.5});
    \draw[dashed,blue,thick] plot[domain=0:3,samples=200] (\x,{erf(-(\x-1.5)*5)/2-5.5});
    \end{scope}
    \begin{scope}[yshift=-1.5cm,xshift=4.5cm,yscale=.5]
    \draw[red,ultra thick,->] (0,-6) -- (3,-6);
    \draw[black, very thick,->] (0,-6) -- (0,-4.5);
    \draw[gray,thick] (0,-5) -- (0.05,-6) -- (3,-6);
    \draw[dashed,blue,thick] (0,-5) -- (0.05,-6) -- (3,-6);
    \end{scope}

    \draw [decorate,decoration={brace,amplitude=10pt},xshift=-4pt,yshift=0pt]
(4.1,-4.6) -- (4.1,-1.7);

    \draw (4.2,-1.4) node[right]{Thermodynamic limit:};
    
    \draw (7,-2.2) node{(a)};
    \draw (7,-3.2) node{(b)};
    \draw (7,-4.2) node{(c)};
    
    \end{tikzpicture}
    \caption{Comparison of the three potential cases in the thermodynamic limit, \ie that the respective subset for a given probe $f(\epsilon)$ may 
    (a) form a phase with sharp transition, (b) have a crossover with soft transition, or (c) form a lower-dimensional critical surface. We plot a probe function $f(\epsilon)$, where $\epsilon$ describes the distance from a given surface.}
    \label{fig:transition}
\end{figure}
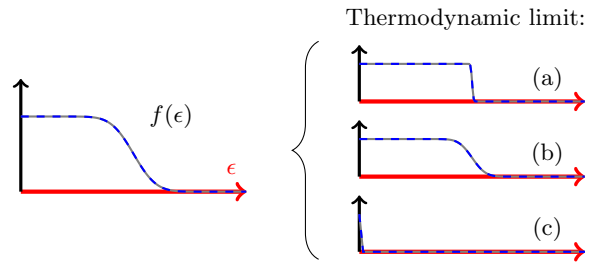

While the focus of this paper lies on the study of Hilbert space fragmentation in a perturbative regime near an exactly fragmented model, the question regarding the nature of transition between different regimes is more general. In many cases, it is natural to define probes, \ie functions $f(\epsilon)\equiv f(\hat{H}_\epsilon)$, which are expected to characterize the respective properties in the different regimes. For finite systems, one generally expects a smooth or even analytical transition between different regimes, but it is a natural question to analyze how this transition behaves in the thermodynamic limit. Generally, we can distinguish three cases (see Fig.~\ref{fig:transition}):   (a)~there are separate phases, that is $f$ is nonanalytical at some critical value~$\epsilon^*$; (b)~there is a crossover, that is $f$ stays analytical or smooth in the transitional regime; or (c)~there is a critical, lower-dimensional surface, \ie the probe function is discontinuous exactly at $\epsilon=0$ in the thermodynamic limit. Note that we focus on a given probe $f$ relative to a single parameter $\epsilon$. We can consider different probe functions and different parametrizations $\hat{H}_{\epsilon}$ to explore higher dimensional regions.

\subsection{Quantifying fragmentation}

Previous works have looked at the interplay between different types of fragmentation and the possibility of localized dynamics in various one-dimensional models \cite{PhysRevB.101.214205, frey_hilbert_2022}, concluding that they seem to be independent. One may find strong fragmentation along with localized or delocalized dynamics. The notion of localization in this context differs from many-body localization and is concerned with the mobility of charges as indicated by active and inactive lattice sites. In a single-particle problem this kind of localization may be detected by the standard IPR defined for single-particle wavefunctions. For a many-body system a more general IPR may be defined that is sensitive to this notion of localization.

Parametrized MBL models typically have a trivial limit in which eigenstates are exactly local product states. This makes an IPR defined with respect to this product basis a suitable measure of many-body localization. However, the ideal limit of a fragmented model only constraints eigenstates to blocks with respect to a local product basis. Typical blocks may grow in size with increasing system size, leading to a vanishing IPR in the thermodynamic limit. In order to circumvent this issue, we define an inverse participation ratio with respect to the ideal block structure. We then show that this quantity captures the crossover or transition between fragmentation and a quantum chaotic regime when considering finite perturbations by comparison to other indicators such as entanglement entropy and level-spacing statistics: 
The \emph{eigenstate entanglement entropy} studies the bipartite entanglement entropy averaged over all energy eigenstates, which for general quantum chaotic systems is expected to satisfy a volume law \cite{PhysRevX.8.021026,bianchi2021volume,PhysRevE.97.012140,HUANG2019594}. The leading-order asymptotics should be independent of model parameters within a thermal regime, since it is determined by the tensor product structure of the Hilbert space and conserved quantities alone. 
The \emph{level-spacing statistics} studies the normalized distance between different energy eigenvalues (after taking physical symmetry sectors into account) and is expected to follow the predictions of the Gaussian orthogonal ensemble for quantum chaotic systems~\cite{PhysRevLett.52.1}.


\section{Inverse Participation Ratios for fragmented systems}

In this section we discuss different definitions of the inverse participation ratio and their applicability. We first consider the older definitions of single-particle, one-particle-reduced, and product basis IPR, and the context in which they arise naturally. We then elucidate potential issues when applied to fragmented systems and present the block IPR as a suitable alternative in this case.

\subsection{Traditional inverse participation ratios (IPRs)}

The state of a single quantum particle on a lattice defined by the set of points $\{x\}$ is characterized by the normalized wave-function $\psi(x)$ with $\sum_x |\psi(x)|^2 = 1$. One may define the single-particle IPR via

\begin{equation}
    \mathrm{IPR}_\mathrm{sp} = \sum_x |{\psi(x)}|^4 .
\end{equation}

It can easily be verified that $1/L \leq \mathrm{IPR}_\mathrm{sp} \leq 1$, where $L$ is the number of lattice sites. $\mathrm{IPR}_\mathrm{sp}$ is maximal when $\psi$ only has finite weight on exactly one lattice site, i.e., it represents a fully localized state, and minimal when it has equal weight on all lattice sites, representing a fully delocalized state. 

The generalization of  $\mathrm{IPR}_\mathrm{sp}$ to $N > 1$ particles on the lattice $\{x\}$ is not obvious, since the state is now a wave-function $\psi(x_1,\dots,x_N)$ on $\{x\}^N$. Considering a fermionic many-body system, one possible approach \cite{bera_many-body_2015} is to consider the one-particle reduced density matrix $\rho$, which in the language of second quantization may be expressed as

\begin{equation}
\rho_{x x'}=\bra{\psi} \hat{c}_x^{\dagger} \hat{c}_{x'} \ket{ \psi} ,
\end{equation}

\noindent with $\hat{c}^\dagger$ creation and $\hat{c}$ annihilation operators. $\rho$ can then be diagonalized to give eigenvectors $\phi_\alpha$ with eigenvalues $0 \leq n_\alpha \leq 1$ with $\sum_\alpha n_\alpha = N$, 

\begin{equation}
\rho \phi_\alpha = n_\alpha  \phi_\alpha .
\end{equation}

Since $\phi_{\alpha,x}$ is a unitary matrix, one may choose $\hat{\Tilde{c}}_\alpha = \sum_x \phi_{\alpha,x} \hat{c}_x$ as a new fermionic basis. The many-body state $\ket{\psi}$ can be characterized as one where the single-particle orbitals $\Tilde{c}_\alpha$ each have occupancy $n_\alpha$. A natural generalization of  $\mathrm{IPR}_\mathrm{sp}$ is therefore the one-particle reduced IPR defined as the weighted average of single-particle IPRs of the orbitals $\Tilde{c}_\alpha$,

\begin{equation}\label{eq:IPR_opr}
     \mathrm{IPR}_\mathrm{opr} = \frac{1}{N} \sum_\alpha n_\alpha \sum_x | \phi_{\alpha,x} |^4 .
\end{equation}

This definition may be used to quantify many-body localization in finite systems.

A different approach is to pick an orthonormal basis $\{\ket{b}\}$ of the many-body Fock space $\mathcal{H}$ and define the many-body IPR for a state $\ket{\psi} = \sum_b \alpha_b \ket{b}$ with $\sum_b |\alpha_b|^2 = 1$ as 

\begin{equation}
     \mathrm{IPR}_\mathrm{basis} = \sum_b | \alpha_b |^4   .
\end{equation}
What constitutes a sensible choice of $\{\ket{b}\}$ depends on the system. Generally we can say that $1/\mathrm{dim}(\mathcal{H}) \leq \mathrm{IPR}_\mathrm{basis} \leq 1$, where the lower bound is realized if the state has equal weight over the entire basis. This may be the case for Haar-random states, in which case the IPR is independent of the choice of $\{\ket{b}\}$ unless fine tuned. Typically one might choose a local product basis in the context of MBL, since thermal states should look highly random when expressed in terms of local product states. A many-body localized state, however, may have finite weight on a significantly smaller number of local product states. We will use the "standard model" of MBL as an example, which can be expressed in the spin language or fermionic language respectively as

\begin{equation}\label{eq:MBL-Hamiltonian}
\begin{aligned}
    \hat{H}_\mathrm{MBL} = \sum_{x=1}^L \, & h_x \hat{\sigma}^z_x + J \, \hat{\sigma}^z_{x} \hat{\sigma}^z_{x+1}   \\
    &+ J \left( \hat{\sigma}^x_{x} \hat{\sigma}^x_{x+1} + \hat{\sigma}^y_{x} \hat{\sigma}^y_{x+1} \right) \\
    = \sum_{x=1}^L & \left[ \vphantom{\left( \hat{c}^\dagger_{x+1} \right)} 2 h_x \hat{n}_x + 4J \, \hat{n}_{x} \hat{n}_{x+1} - h_x \right.  \\ 
    & \left. + \, 2J \left( \hat{c}^\dagger_x \hat{c}_{x+1} + \hat{c}_x \hat{c}^\dagger_{x+1} \right)  \right] + J( L-4N ) \;,
\end{aligned}    
\end{equation} 
where $\hat{n}_x = \hat{c}^\dagger_x \hat{c}_x$ is the particle number at site $x$, $N$ the total number of particles, and $h_x$ are real random numbers. The limit $J \to 0$ is trivially localized, since the eigenstates of $\hat{H}_\mathrm{MBL}$ become bit strings in the fermionic language. It is therefore natural to choose the set of bit strings with a fixed total number of occupied sites $N$ as our basis $\{\ket{b}\}$ and to use the corresponding $\mathrm{IPR}_\mathrm{basis}$ as a probe for MBL when $\epsilon>0$. We note that a different choice of local product basis can yield very different results even if the state is localized. For example, the state $\ket{\rightarrow\rightarrow\dots\rightarrow}$, here expressed in the spin language, has minimal $\mathrm{IPR}_\mathrm{basis} = 1/{2^L}$ with respect to the basis $\{\ket{\muparrow/\mdownarrow\dots\muparrow/\mdownarrow}\}$. Unlike entanglement entropy, using IPR as a probe of ergodicity requires some knowledge of the non-ergodic limit of the model.

\subsection{Definition: Block inverse participation ratio}

Finally, we consider fragmented systems. The ideal, non-ergodic limit of a fragmented model is characterized by a Hamiltonian that is block diagonal (typically) with respect to a local product basis after resolving all conventional symmetries. This suggests that the $\mathrm{IPR}_\mathrm{basis}$ defined with respect to said product basis is a good measure for the extend of fragmentation when considering finite perturbations away from the exact limit. However, there are two potential issues: The size of typical blocks may grow with increasing system size, such that $\mathrm{IPR}_\mathrm{basis} \to 0$ in the thermodynamic limit both in the fragmented regime and the thermal regime. Second, $\mathrm{IPR}_\mathrm{basis}$ is expected to be sensitive to additional perturbations such as local disorder, which either do not or only weakly affect the level to which eigenstates respect the ideal block structure. We therefore define a block IPR with respect to the set $\mathcal{B}$ of ideal blocks of the exactly fragmented limit

\begin{equation}\label{eq:bIPR}
     \mathrm{IPR}_\mathrm{block} \left( \ket{\psi} \right) = \sum_{b \in \mathcal{B}} \lVert \hat{P}_b \ket{\psi} \rVert ^4 \quad ,
\end{equation}
with $\hat{P}_b $ the projector onto block $b$. This quantity ranges between $1/\#\mathcal{B} \leq \mathrm{IPR}_\mathrm{block} \leq 1$, where $\#\mathcal{B}$ is the cardinality of $\mathcal{B}$. Unlike $\mathrm{IPR}_\mathrm{basis}$, it is expected to approach 1 in the fragmented limit independently of system size $L$, whereas the lower limit approaches $1/\#\mathcal{B} \to 0$ as $L\to\infty$.


\section{Three probes of fragmentation breaking}

In order to study the emergence and dissolution of the fragmented block structure numerically, we consider three different probes. Two of these probes, the statistics of level-spacing and bipartite entanglement entropy, are standard indicators of thermal to nonthermal transitions. 

The concept of level-spacing statistics has its origin in random-matrix theory and we expect the spectrum of a generic real, symmetric Hamiltonian to match the statistics of the Gaussian orthogonal ensemble. On the other hand, conventionally integrable models should satisfy Poisson statistics. An emergent integrability such as in many-body localized systems also yields Poisson statistics. In the case of fragmentation, the particular value of the level-spacing parameter may not be universal, hence we take any significant deviation from GOE as an indicator for the emerging block diagonalization. 

Real-space entanglement entropy of eigenstates is expected to satisfy a volume law for thermal systems, while many-body localization implies an area law. Fragmentation predicts a reduction in entanglement entropy relative to the thermal value due to the emergent constraints on the eigenstates.

While level statistics and entanglement entropy are quite universal probes and basis independent, inverse participation ratios are typically basis dependent and specific to the problem. We study the block IPR introduced in Eq.~(\ref{eq:bIPR}) and compare it to the other numerical probes in order to establish its validity as an indicator of the transition between thermal and fragmented regimes.

\subsection{Block inverse participation ratio}
Our first probe is the inverse participation ratio with respect to the fragmented block structure $\mathrm{IPR}_\mathrm{block}$ [Eq.~(\ref{eq:bIPR})]. Given a family of Hamiltonians $\hat{H}(\boldsymbol{\eta})$ parametrized by $\boldsymbol{\eta} = \left(\eta_i \right)_{i=1,\dots,m}$ that shows fragmentation in the limit $\eta_i \to \infty$, we calculate all (normalized) energy eigenvectors $\ket{E_m^{(N,k)}(\boldsymbol{\eta})}$ for fixed particle number $N$ and momentum $k$ and compute the averaged IPR with respect to the block structure found in said limit $\eta_i \to \infty$,
\begin{equation}
     \mathrm{IPR}_{\mathrm{block},i}^{(N,k)} (\boldsymbol{\eta})
     = \frac{1}{\mathcal{N}(N,k)} \sum_m \sum_{b \in \mathcal{B}^i } \lVert \hat{P}_b \ket{E_m^{(N,k)}(\boldsymbol{\eta})} \rVert ^4   ,
\end{equation}
where $\mathcal{N}(N,k)$ is the number of states with particle number $N$ and momentum $k$, and $\mathcal{B}^i$ is the set of $M$ blocks, \ie disconnected subspaces $\left( \mathcal{H}^i_j \right)_{j=1,\dots,M}$ of the Hilbert space $\mathcal{H} = \bigoplus_{j=1}^M \mathcal{H}^i_j $, found in this limit. The $\mathcal{H}^i_{j}$ may be identified as the connected components of the adjacency matrix $\hat{H}^{i}_\infty$, the latter being obtained as the effective Hamiltonian that is the limit $\mathrm{lim}_{\eta_i \to \infty} \hat{H}(\boldsymbol{\eta})$.
In the limit of exact fragmentation the eigenvectors can be chosen such that $\ket{E_m^{(N,k)}(\boldsymbol{\eta})} \in \mathcal{H}^i_{j(n)}$. One would therefore naively expect $\lim_{\eta_i \to \infty} \mathrm{IPR}_{\mathrm{block},i}^{(N,k)} (\boldsymbol{\eta}) = 1$. However, the block structure of  $\hat{H}^{i}_\infty$ is generally not invariant under exact symmetries of the model, such as translation by one site $\hat{T}$, leading to exactly degenerate blocks and possibly a discontinuity in $ \mathrm{IPR}_{\mathrm{block},i}^{(N,k)} (\boldsymbol{\eta})$ in the limit. In order to address this issue, we use perturbation theory around the exactly fragmented limit (see Sec.\ref{sec:Schrieffer}).

\subsection{Level-spacing}\label{sec:LevelSpacing}

The second probe is the statistics of level-spacing. Denoting the $l$th ordered eigenvalue of $\hat{H}$ by $E_l$, one may define the level-spacing parameter $r_l~=~\mathrm{min}\{\delta^{(l)},\delta^{({l+1})}\}/\mathrm{max}\{\delta^{(l)},\delta^{({l+1})}\}$, where $\delta^{(l)}=E_{l+1}-E_l$. This quantity is independent of the overall density of states. Averaging over the index $l$ in the middle of the spectrum yields $\bar{r}$ with a value of $\bar{r}_\mathrm{GOE} \approx 0.5307$ if the system is ergodic and $\bar{r}_\mathrm{Poisson} \approx 0.386$ if the system is localized or otherwise integrable. Here it is assumed that degeneracies due to symmetries have been resolved. It has been shown that unresolved discrete symmetries result in GOE-like statistics but with a value of $\bar{r}$ that lies in between  $\bar{r}_\mathrm{Poisson}$ and $\bar{r}_\mathrm{GOE}$~\cite{PhysRevX.12.011006,PhysRevLett.110.084101,de_marco_level_2022}. Random matrices comprised of $m$ independent and equal-sized symmetry blocks show level-spacing statistics that interpolates between  GOE${}_{m=1}$=GOE and GOE${}_{m \to \infty}$=Poisson.
This will be relevant when we consider the half-filling sector and is also the reason why we do not consider $k=0$ or $k=L/2$. Since we do not resolve particle-hole symmetry in the half-filling case, we have $m=2$ and therefore expect a value of $\bar{r}_{\mathrm{GOE}_{m=2}} = 0.423$ \cite{de_marco_level_2022}.

\subsection{Eigenstate bipartite entanglement entropy}

Our final probe is the bipartite entanglement entropy $S_A(\ket{\psi})$ of a pure state $\ket{\psi}$ with respect to a Hilbert space decomposition $\mathcal{H}=\mathcal{H}_A\otimes\mathcal{H}_B$, defined as
\begin{align}
 \hspace{-3mm} S_A(\ket{\psi})=-\Tr \rho_A\log\rho_A\quad\text{with}\quad \rho_A=\Tr_B\ket{\psi}\bra{\psi}.
\end{align}
It is well known that this quantity is symmetric under exchanging the subsystems $A\leftrightarrow B$. The maximal entanglement entropy is given by $S_{\max}=\min(\log(\dim\mathcal{H}_A),\log(\dim\mathcal{H}_B))$. Consequently, for a Hilbert space constructed as a tensor product of local Hilbert spaces of dimension $d$ associated to individual lattice sites, the maximal entanglement entropy is given by $S_{\max}=\min(V_A\log{d},V_B\log{d})$, where $V_A$ and $V_B$ represent the number of lattice sites in the respective subsystems (out of a total of $V=V_A + V_B$  lattice sites).

A natural probe of a Hamiltonian $\hat{H}$ with an orthonormal basis of eigenvectors $\ket{E_i}$ is the average eigenstate entanglement entropy~\cite{PhysRevLett.111.127201,Bauer_2013,vidmar2017entanglement} 
\begin{align}
    \braket{S_A}_{\mathcal{J}}=\frac{1}{d_{\mathcal{J}}}\sum_{i\in\mathcal{J}}S_A(\ket{E_i})\,,
\end{align}
where $\mathcal{J}$ is an index set, potentially only including a subset of the eigenstates, \eg in a symmetry sector. An additional subtlety arises if the Hamiltonian has degeneracies, as the entanglement entropy (and its average) will depend on how the $\ket{E_i}$ are chosen in these degenerate subspaces. 

For a general quantum chaotic model, one expects that the eigenstate entanglement entropy satisfy a volume law $\braket{S_A}_{\mathcal{J}}=a(n)V_A+o(V)$, where $V_A$ is the volume of the smaller subsystem in a bipartition and the constant $a(n)$ typically only depends on Hilbert space structure and particle density $n$ per site. For a random pure state in Hilbert space with particle density $n$, the expected bipartite entanglement entropy was computed up to constant order~\cite{bianchi2021volume} as 
\begin{align}\label{eq: volume law}
    \braket{S_A}^0_n=a_0(n)V_A+b_0(n,f)\sqrt{V_A}+c_0(n,f)+o(1)\,,
\end{align}
where $f=\frac{V_A}{V}$ describes the subsystem fraction of the smaller subsystem. While $a_0(n)$ appears to be universal with $a(n)=a_0(n)$ for quantum chaotic models, \ie the analytical calculation for a Haar random pure state coincides with the finite-size scaling analysis of the eigenstate entanglement entropy for general quantum chaotic models, the next order coefficients $b(n,f)$ and $c(n,f)$ appear to be mildly model dependent. Still, the analytical prediction for a Haar random pure state often gives a good approximation of the model-dependent thermodynamic value.

Integrable models and in particular quadratic fermionic models~\cite{vidmar2017entanglement,vidmar2018volume,hackl2019average,leblond2019entanglement} have been found to also satisfy such a volume law, but with a model-dependent coefficient $a=a(n,f)$, which also depends explicitly on the subsystem fraction $f$, such that $a(n,f)<a_0(n)$ when compared to the entanglement entropy of a random pure state. We expect a similar discrepancy in the fragmented regime of Hamiltonians that show Hilbert space fragmentation. As the difference $a_0(n)-a(n,f)$ is maximal at $f=\frac{1}{2}$, we evaluate the number
\begin{align}\label{eq:smallS}
    s_A=\frac{\braket{S_A}_{\mathcal{J}}(L/2)}{\braket{S_A}^0_n(L/2)}\simeq \frac{a(n,f)}{a_0(n)}\,,
\end{align}
\ie we compare the average eigenstate entanglement entropy to the average entanglement entropy of a random pure state at half of the system size. We expect $s_A \to 1$ in the thermodynamic limit within the quantum chaotic regime, while $s_A < 1$ in integrable, MBL or fragmented regimes. For finite systems, we expect a significant quantitative change when crossing between a thermal and a non-thermal regime, such as a drop in $s_A$ at the onset of fragmentation. We will use this drop as a numerical indicator to distinguish between different regimes or phases when the respective thermodynamic limits are not satisfied due to insufficient system size.


\section{Application: Extended Fermi-Hubbard model} \label{sec:extendedFermi}

We consider the $t$-$V_1$-$V_2$ spinless fermionic chain with periodic boundary conditions
\begin{equation}\label{eq:xfHubbard}
\begin{aligned}
    \hat{H}=&-t\sum_x (\hat{c}^\dagger_{x+1}\hat{c}_x+\mathrm{H.c.}) \\ 
    &+ V_1\sum_{x}\hat{n}_x\hat{n}_{x+1} + V_2\sum_{x}\hat{n}_x\hat{n}_{x+2} \,.
\end{aligned}
\end{equation}
This model has been studied as a genuine fermionic model in \cite{frey_hilbert_2022}, but also in its closely related hard-core boson form~\cite{cazalilla2011one,leblond2021universality,bianchi2021volume}. The important limit $V_1\to \infty$ and $V_2=0$ was explicitly worked out in~\cite{Dias_2000}. We will henceforth set $t=1$.

In this limit ($V_1 \to \infty$, $V_2 = 0$) the hopping term is locally constrained as
\begin{equation}
    \hat{H}^{(1)}_\infty = - \sum_x \hat{P}^{(1)}_x (\hat{c}^\dagger_{x+1}\hat{c}_x+\mathrm{H.c.})\hat{P}^{(1)}_x \, ,
\end{equation}
with the local projector $\hat{P}^{(1)}_x = 1-(\hat{n}_{x+2}-\hat{n}_{x-1})^2$. In the limit $V_2 \to \infty$ and $V_1 = 0$, the hopping term is instead constrained as
\begin{equation}
    \hat{H}^{(2)}_\infty = - \sum_x \hat{P}^{(2)}_x (\hat{c}^\dagger_{x+1}\hat{c}_x+\mathrm{H.c.})\hat{P}^{(2)}_x 
\end{equation}
with the local projector $\hat{P}^{(2)}_x = 1+\frac{1}{4}[(\hat{n}_{x-2}+\hat{n}_{x+2}-\hat{n}_{x-1}-\hat{n}_{x+3})^4 - 5 (\hat{n}_{x-2}+\hat{n}_{x+2}-\hat{n}_{x-1}-\hat{n}_{x+3})^2 ]$. In the limit where $V_1\to \infty$ and $V_2\to\infty$ independently, both constrains act simultaneously. Note that $\hat{P}_x^{(1)}$ and $\hat{P}_x^{(2)}$ commute.

The relevant symmetries of this model are conservation of particle number $\hat{N}$, translation symmetry $\hat{T}$, parity (reflection) $\hat{P}$ and, in the case of half-filling, particle-hole symmetry $\hat{S}$. We will exploit translation symmetry by considering different momentum sectors separately. This allows us to study larger system sizes. In the following, momentum $k$ refers to the eigenspace of the translation operator $\hat{T}$ with eigenvalue $\mathrm{exp}(-i \, 2 \pi k / L )$. We generally find no significant $k$ dependence in our results apart from the effect of discrete symmetries on level statistic outlined above. Hence we present results for a single momentum sector $k=1$.

\subsection{Inverse participation ratios}

In this section we apply two different definitions of IPR to the extended Fermi-Hubbard model (\ref{eq:xfHubbard}). First, we demonstrate how the one-particle reduced IPR Eq.~(\ref{eq:IPR_opr}) can be used to show a tendency towards real-space localized dynamics when combining nearest- and next-nearest-neighbor interactions. Second, we use our block IPR Eq.~(\ref{eq:bIPR}) to quantify the onset of fragmentation when considering the strongly correlated limits. 

\subsubsection{Real-space localization transition within the fragmented regime}

Previous study on the model (\ref{eq:xfHubbard}) has shown that fragmentation due to large $V_1$ or large $V_2$ does not result in localized dynamics, while a combination of large $V_1$ and $V_2$ does \cite{frey_hilbert_2022}. Typical states will be characterized by dynamical bubbles that are disconnected from each other by inactive regions. Eigenstates are therefore expected to show localization as indicated by the one-particle reduced IPR Eq.~(\ref{eq:IPR_opr}). However, translation symmetry may mask this transition. {We therefore add a weak local disorder term $W \sum_x h_x \hat{n}_x$ to break the degeneracy and ensure that numerical diagonalization gives the localized eigenstates whenever possible. We choose uniformly random distributed $h_x \in [-1,1]$ and $W=10^{-2}$, which is too weak to induce MBL~\cite{de2019dynamics}.
Figure~\ref{fig:IPRopr} shows the $V_2$ dependence of $\mathrm{IPR_{opr}}$ for fixed values of $0 \leq V_1 \leq 10^3$ for $L=14$. Taking $V_1 \to \infty$ for $V_2$ held constant results in an increased average $\mathrm{IPR_{opr}}$ due to the emergence of frozen states (indicated by directions of red and green arrows). Eigenstates associated with large blocks remain delocalized at $V_2=0$ and hence the increase in $\mathrm{IPR_{opr}}$ is relatively small (red arrow). This is the motivation behind defining a block IPR that fully displays the fragmented limit. Likewise, taking $V_2 \to \infty$ for $V_1=0$ results in an increased average $\mathrm{IPR_{opr}}$ due to the emergence of frozen states (indicated by purple arrow). Typical states remain delocalized as shown in \cite{frey_hilbert_2022}, but some states associated with large blocks exhibit localization and contribute to this change in $\mathrm{IPR_{opr}}$. 
Crucially we find a much stronger $V_2$ dependence for finite, but relatively small, $V_1$. In particular, for $V_1=10$ (highlighted in red) we find that $\mathrm{IPR_{opr}}$ rapidly approaches the maximum value found at $V_1,V_2 \to \infty$ when increasing $V_2$. This is further demonstration that the dynamics of this system become localized when adding next-nearest-neighbor interactions to the Fermi-Hubbard model with finite nearest-neighbor interactions. The tendency to localize becomes strong when nearest-neighbor interactions approach the threshold $V_1 \sim 10^1$ where our numerical probes indicate the onset of fragmentation (see below).

\begin{figure}[t]
    \centering
    \includegraphics[width=8.5cm]{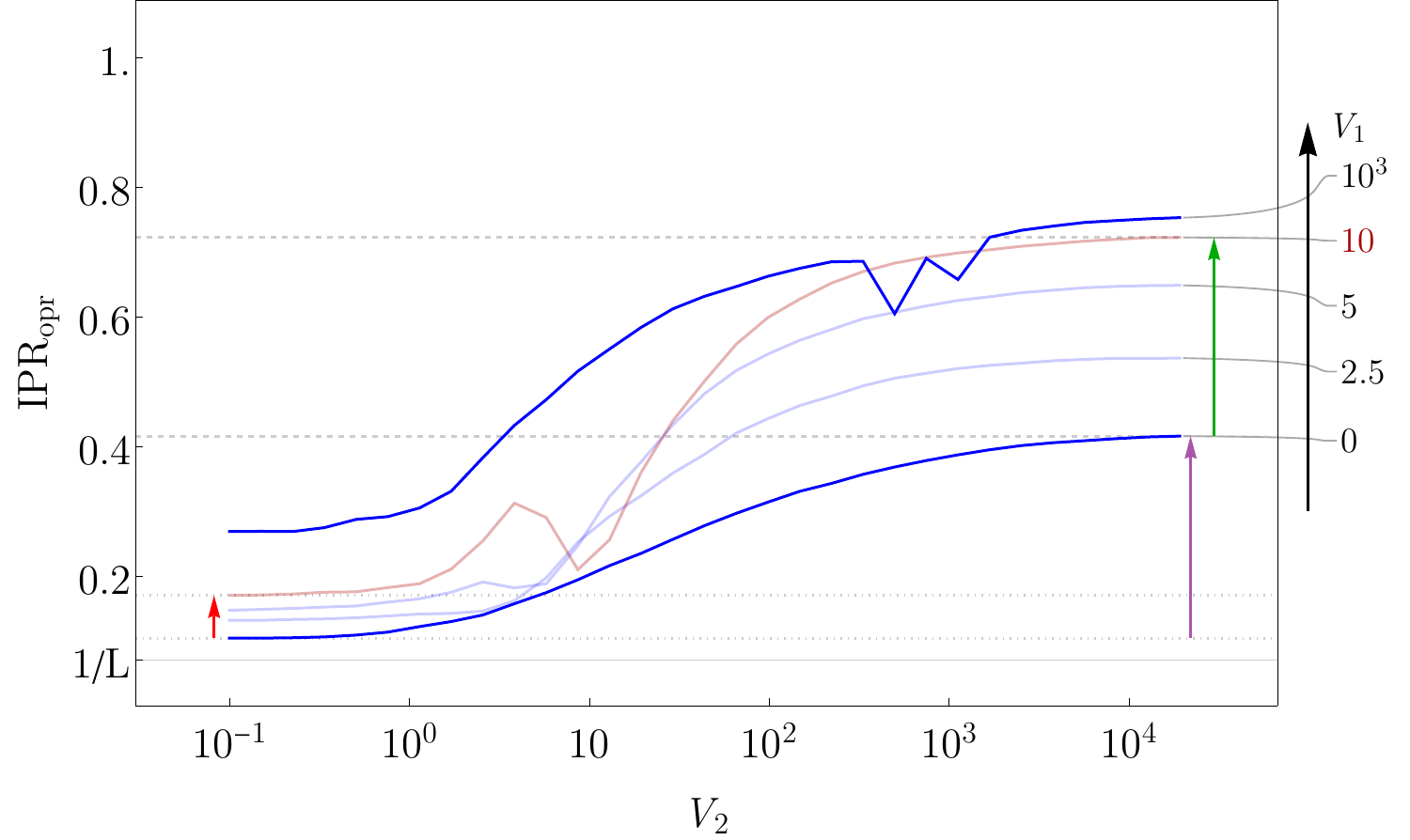}
    \caption{One-particle reduced IPR for $L=14$ and different fixed values of $V_1$. Red arrow: Increase in IPR due to frozen states emerging as $V_1 \to \infty$ for $V_2 \approx 0$. Purple arrow: Increasing IPR due to frozen states emerging as $V_2 \to \infty$ for $V_1 = 0$. Green arrow: Effect of additional localization of states in large blocks at finite $V_1 \sim 10$ when adding nearest-neighbor interactions $V_2 \gtrsim 10$ (compare to red arrow). Based on our previous analysis of this model~\cite{frey_hilbert_2022} we expect reduced fragmentation around $V_1=V_2$ and $V_1=2V_2$ due to weaker constraints. This is indicated by dips in the one-particle reduced IPR, which are resolved as two separate features for very large $V_1=10^3$.}
    \label{fig:IPRopr}
\end{figure}

\begin{figure*}[t]
    \centering
    \begin{tikzpicture}
        \draw (-6.2,2.6) node{\textbf{(a) Block IPR}};
        \draw (-.4,2.6) node{\textbf{(b) Level-spacing}};
        \draw (5.9,2.6) node{\textbf{(c) Entanglement entropy}};
        
        \draw (-6,0) node{\includegraphics[height=4.6cm]{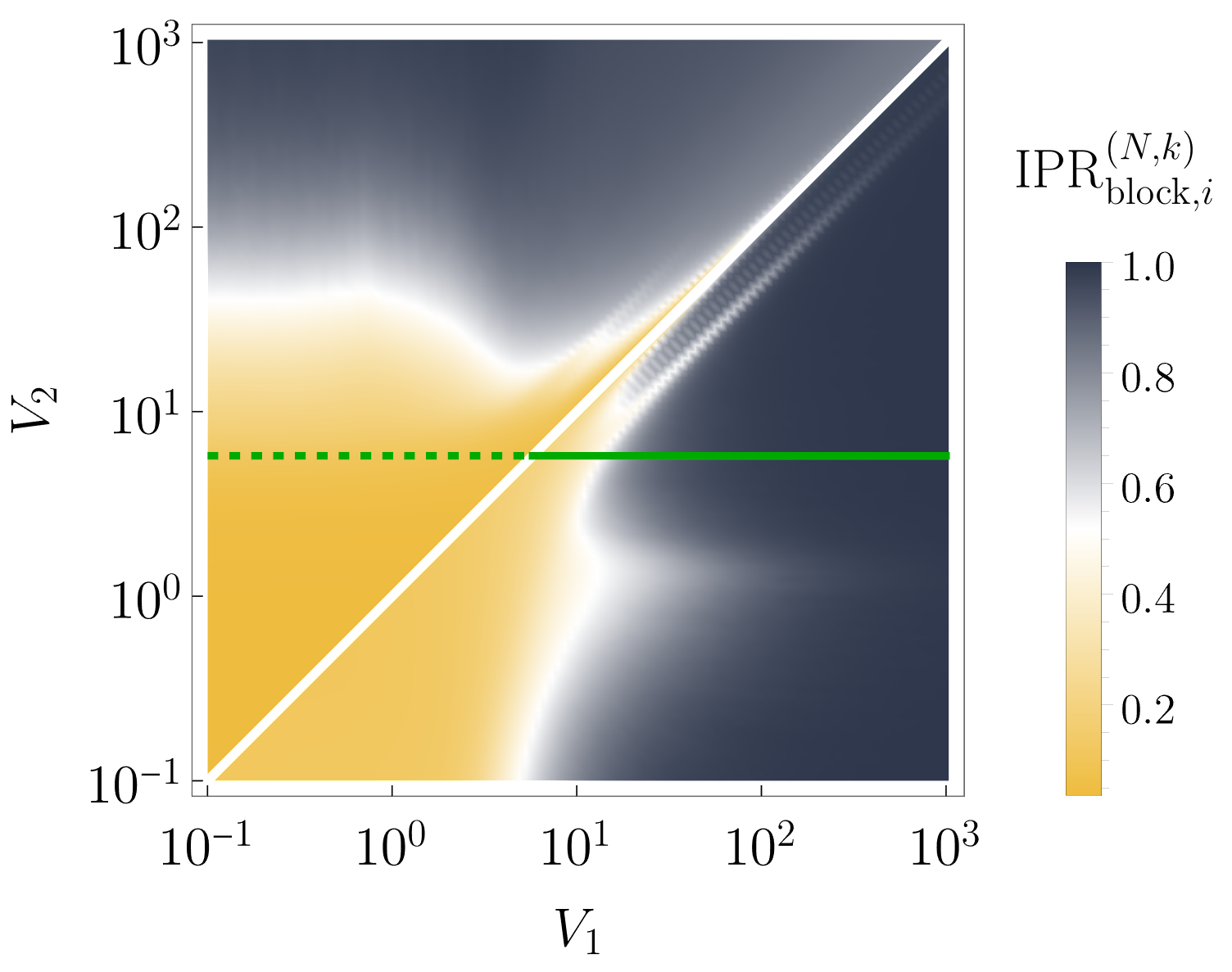}};
        \draw (0,0) node{\includegraphics[height=4.6cm]{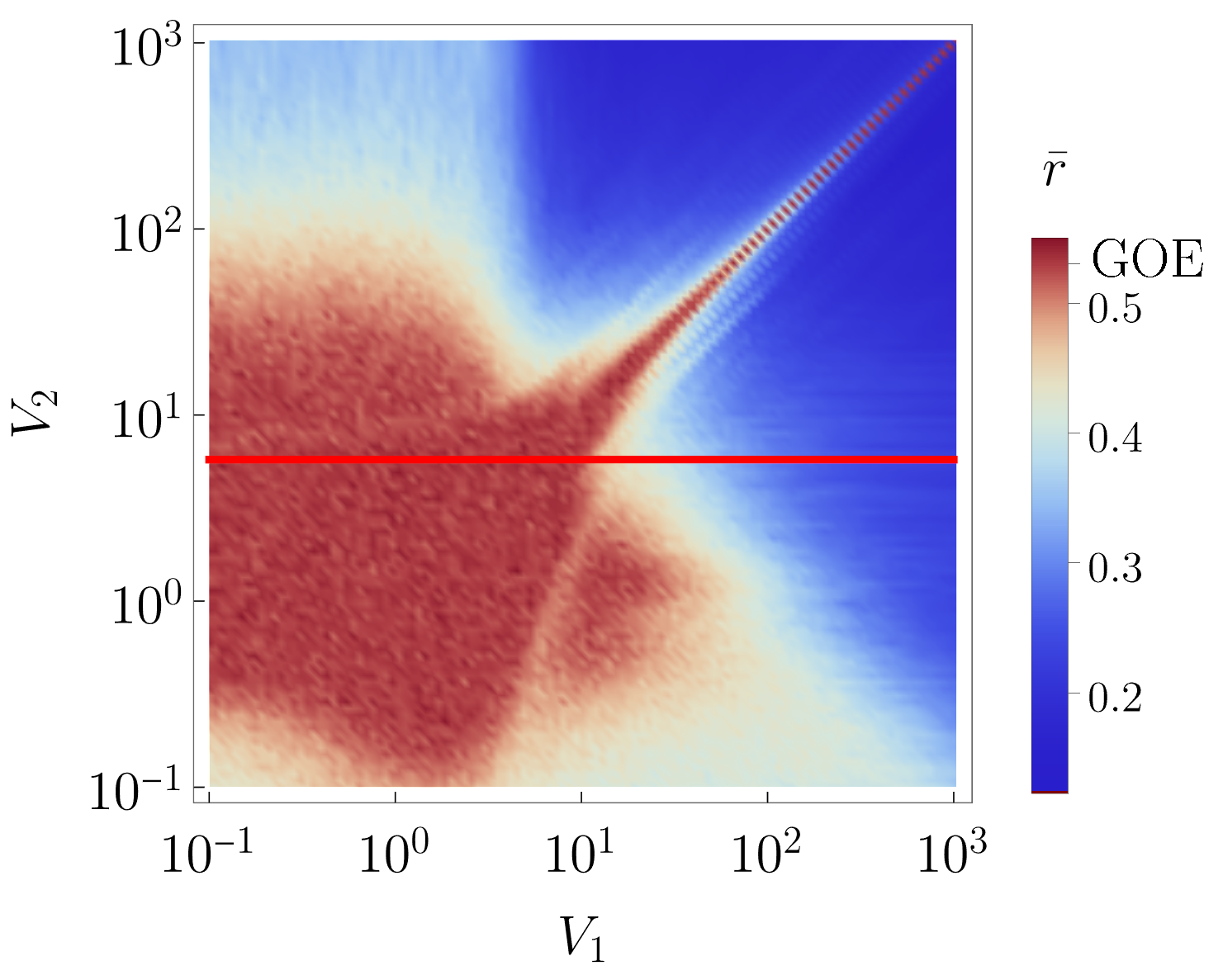}};
        \draw (6,0) node{\includegraphics[height=4.6cm]{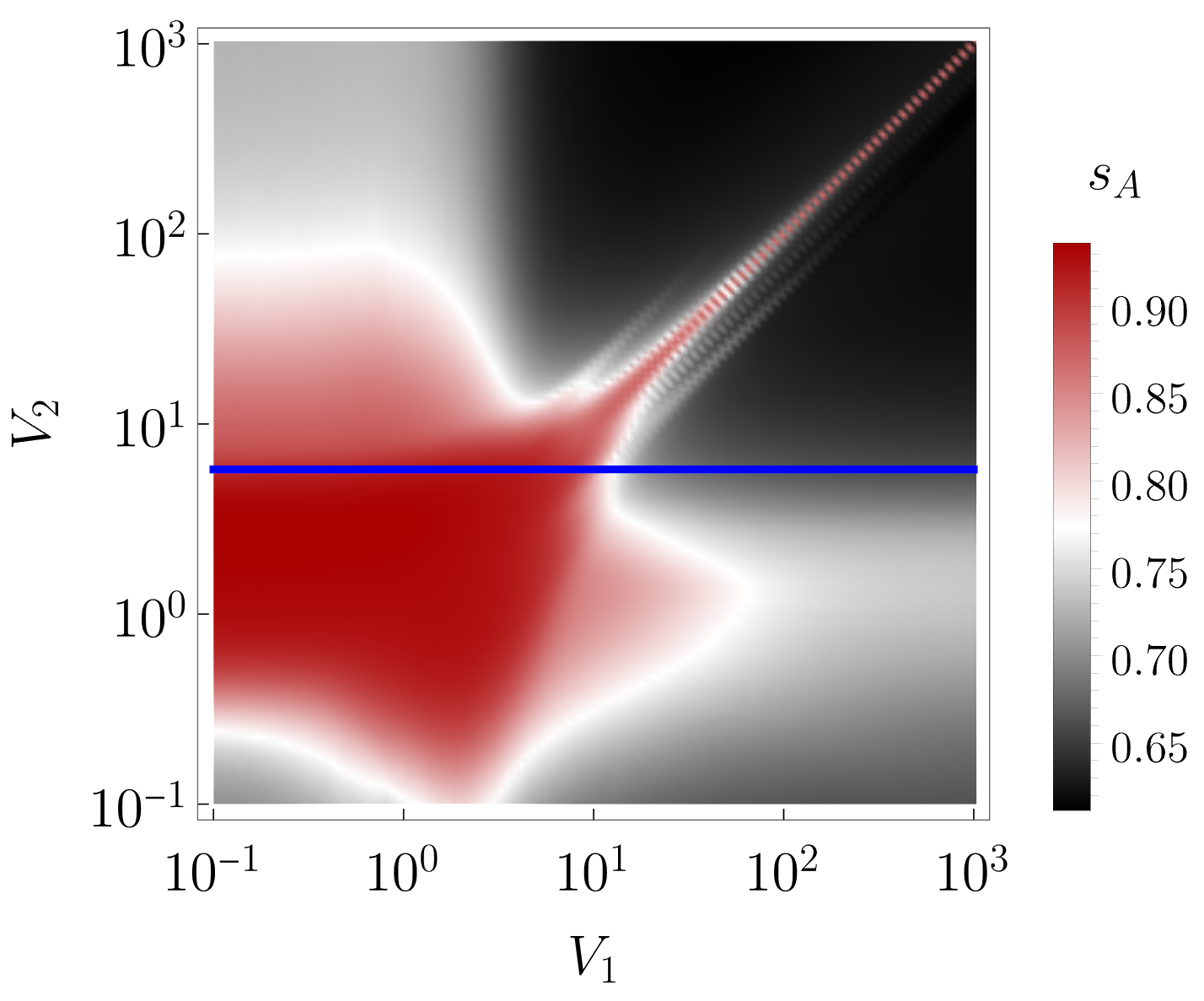}};



        \draw (-7.15,-1.2) node{$i\!=\!1$};
        \draw (-7.7,-.6) node[rotate=90]{$i\!=\!2$};
        
        \begin{scope}[yshift=-5mm]
        \draw (-4,-5) node{\includegraphics[width=7cm]{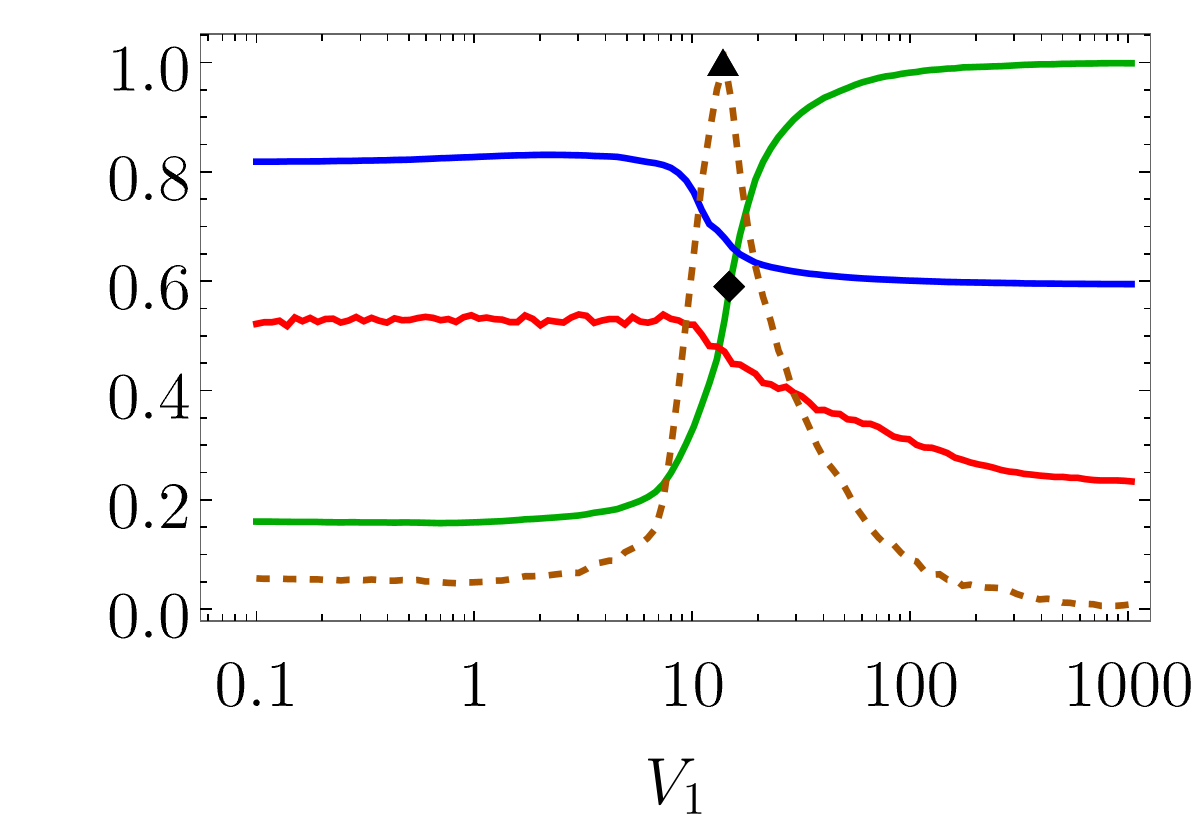}};
        \draw (-3.8,-2.3) node{\textbf{(d) Horizontal cut of transition probes (a)-(c)}};
        \draw (-6.1,-5.2) node[red,right]{$\bar{r}$} (-1.8,-3.4) node[dgreen]{$\mathrm{IPR}^{(N,k)}_{\mathrm{block},1}$} (-6.1,-3.2) node[blue,right]{$s_A$} 
        
        
        (-1.6,-5.03) node[dorange,text width=1.6cm, align=left]{normalized $\mathrm{IPR}^{(N,k)}_{\mathrm{block},1}$ \vskip 0.15cm variance};
        
        \draw (4,-5) node{\includegraphics[width=7cm]{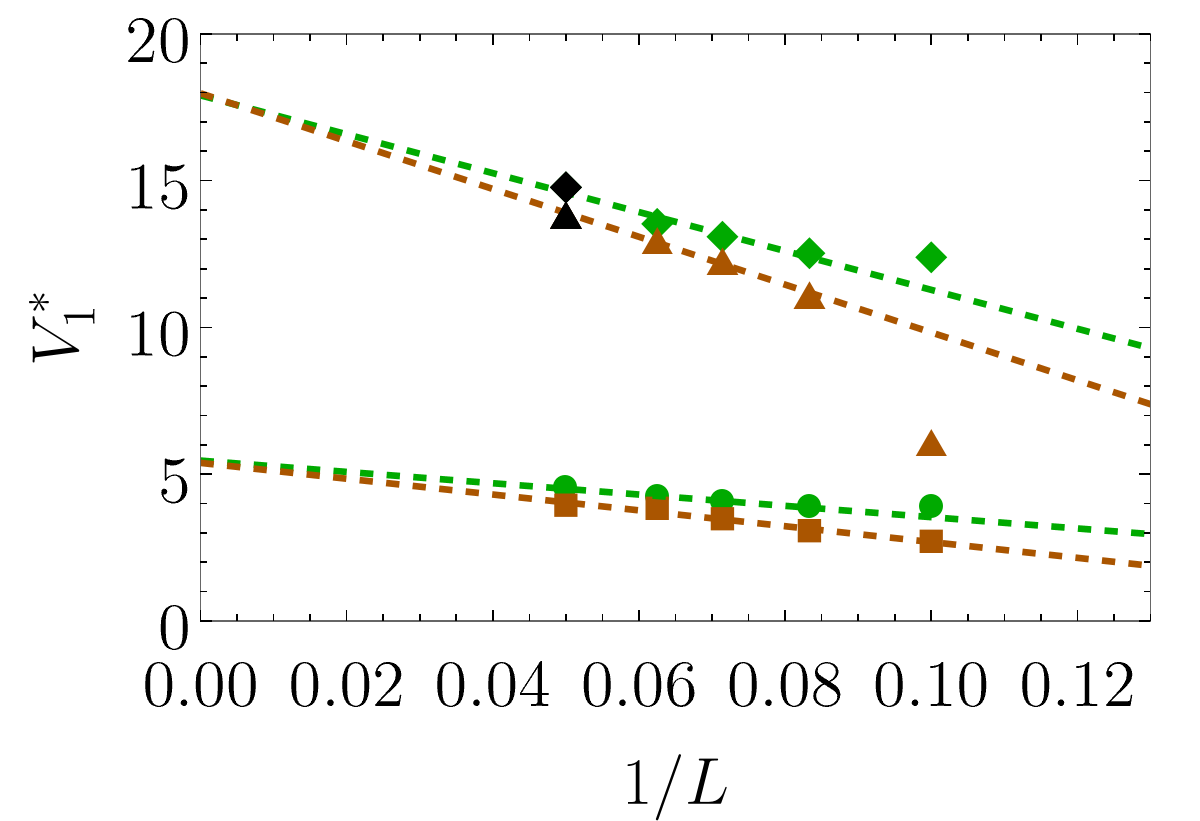}};
        \draw (4.3,-2.3) node{\textbf{(e) Scaling analysis transition point $V^*_1$}};
        \draw (3.3,-5.8) node{$V_2=0$} (4.4,-4.6) node{$V_2=5.77$};
        \draw (7.1,-3.6) node[left,dgreen]{midpoint $\mathrm{IPR}^{(N,k)}_{\mathrm{block},1}$} (7.1,-3.15) node[left,dorange]{peak $\mathrm{IPR}^{(N,k)}_{\mathrm{block},1}$ variance};
        \end{scope}
    \end{tikzpicture}
\caption{Comparison of three different numerical probes for $L=20$, $N=9$ and momentum $k=1$. \textbf{(a)} Block IPR defined with respect to the effective block structure at $V_1 \to \infty$ (lower right half) and the effective block structure at $V_2 \to \infty$ (upper left half). \textbf{(b)} Level-spacing statistics with color scale highlighting GOE to indicate thermal regime. \textbf{(c)} Average eigenstate entanglement entropy normalized to theoretical prediction for quantum chaotic system (compare~\ref{eq:smallS}). \textbf{(d)} Constant-$V_2$ cuts of all three probes corresponding to colored lines in \textbf{(a)-(c)}. The black diamond indicates the midpoint of the jump in average $\mathrm{IPR}_\mathrm{block}$, while the black triangle indicates the position of maximal variance in $\mathrm{IPR}_\mathrm{block}$ across eigenstates.  \textbf{(e)} Scaling analysis of the respective values for $V_1^*$ obtained in \textbf{(d)} across varying system size and for two different values of $V_2$ (fit is based on the four largest values of $L$).}
    \label{fig:3-panel}
\end{figure*}
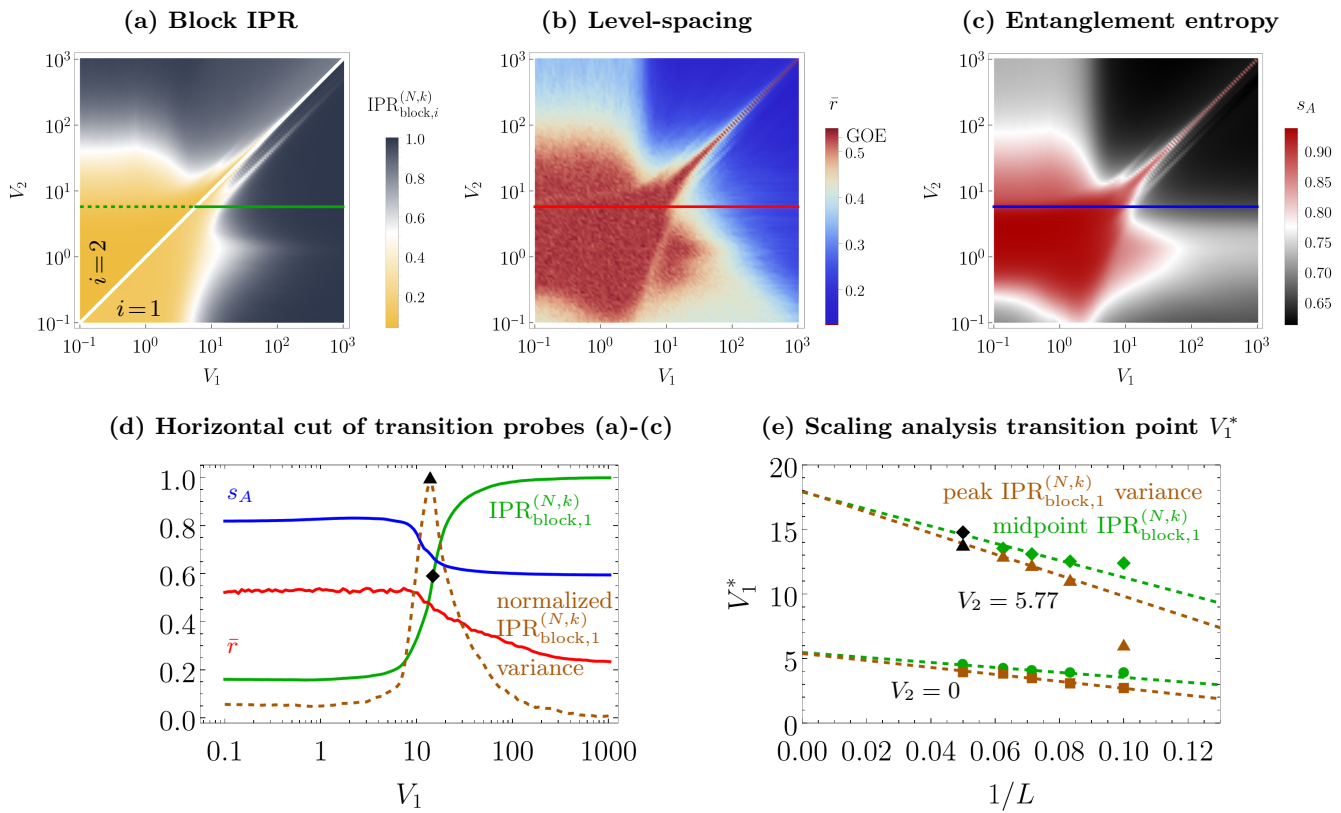

\subsubsection{Block IPR}

In Fig.~\ref{fig:3-panel}(a) we show the average block IPR of exact eigenstates as a function of $V_1$ and $V_2$ at $L=20$, $N=9$ and momentum $k=1$. The lower right triangle displays $\mathrm{IPR}_\mathrm{block}$ computed with respect to the block structure found at $V_1 \to \infty$, while the upper left triangle displays  $\mathrm{IPR}_\mathrm{block}$ computed with respect to the block structure found at $V_2 \to \infty$. We find a distinct region of low $\mathrm{IPR}_\mathrm{block}$ for sufficiently small $V_1, V_2$. It approaches the maximum value of $\mathrm{IPR}_\mathrm{block}$=1 even at moderate values $V_1,V_2 \sim 10^1$.

\subsubsection{Scaling analysis}

Figure~\ref{fig:3-panel}(d) shows block IPR along with level statistics and entanglement entropy (discussed below) taken from a cut along fixed $V_2 = 5.77$ [see colored lines in Fig.~\ref{fig:3-panel}(a)-\ref{fig:3-panel}(c)], along with the normalized variance of $\mathrm{IPR}_\mathrm{block}$. The transition indicated by a sharp increase in $\mathrm{IPR}_\mathrm{block}$ is accompanied by drops in $\bar{r}$ and $s_A$. In addition, the variance of $\mathrm{IPR}_\mathrm{block}$ spikes at the same value of $V_1$. We consider the maximum of this variance, as well as the mid-point of the $\mathrm{IPR}_\mathrm{block}$ itself as indicators of a critical value $V_1^*$. Figure~\ref{fig:3-panel}(e) shows a finite-size scaling analysis based on these two indicators for $V_2 = 5.77$ and for the integrable limit $V_2 = 0$. We see that both quantities converge to the same value as $L \to \infty$ and indicate a finite critical value $V_1^*$. Comparing with Fig.~\ref{fig:transition}, we seem to have ruled out option (c), meaning that $\mathrm{IPR}_\mathrm{block}$ seems to define a finite region around the exactly fragmented limit that matches the non-thermal regime indicated by more conventional quantities like level statistics and entanglement entropy. 

\subsection{Level statistics} 

Figure~\ref{fig:3-panel}(b) shows the level-spacing parameter $\bar{r}$ as a function of $V_1$ and $V_2$. We find that there exists a thermal island at intermediate values of $V_1$ and $V_2$ that is consistent with the island of low $\mathrm{IPR}_\mathrm{block}$ found in Fig.~\ref{fig:3-panel}(a). Since the data is obtained for $L=20$, $N=9$, we see conventional GOE statistics within the thermal region, as opposed to the generalized $\mathrm{GOE}_{m=2}$ observed at half-filling (see Appendix \ref{App:MorePlots}). We also find that the level statistics follow GOE along the diagonal $V_1=V_2$, indicating that in this case the local constraints are sufficiently weak and do not lead to a fragmented block structure. However, there is an additional nonthermal region at small values of $V_2$ that is not captured by the block IPR. This is due to proximity to the exactly integrable case $V_2=0$ and shows that $\mathrm{IPR}_\mathrm{block}$ captures nonergodicity caused by fragmentation specifically.

\subsection{Entanglement entropy}
We evaluate the probe $s_A$ introduced in Eq.~\eqref{eq:smallS} for the fixed momentum sectors $k$ and fixed filling $n=N/L$ at given system size $L$, as shown in Fig.~\ref{fig:3-panel}. Note that $s_A$ is defined as first evaluating the average half-system eigenstate entanglement entropy $\braket{S_A}_{\mathcal{J}}$ with respect to eigenstates in this sector (fixed $k$, $n=N/L$ and $L$) and then normalizing this quantity by dividing the average entanglement entropy $\braket{S_A}^0_n$ of a Haar random state in this filling sector.

For the system under consideration, the leading order coefficient of $\braket{S_A}^0_n$ [compare Eq.~(\ref{eq: volume law})], expected to be universal, is given by~\cite{PhysRevX.8.021026,vidmar2017entanglement,bianchi2021volume}
\begin{align}
    a_0(n)=-n\log(n)-(1-n)\log(1-n) \, .
\end{align}
The finite-size corrections for typical pure states are given by~\cite{bianchi2021volume}
\begin{align}
    b_0(n,f)&=-\sqrt{\frac{n(1-n)}{2\pi}}\log\frac{1-n}{n}\delta_{f,\frac{1}{2}}\,,\\
    c_0(n,f)&=\frac{f+\log(1-f)}{2}-\frac{1}{2}\delta_{f,\frac{1}{2}}\delta_{n,\frac{1}{2}}\,,
\end{align}
which we can use as an approximation for the true model-dependent values. This allows us to compute $\braket{S_A}^0_n$ up to constant order, which we use then to evaluate $s_A$. For $L=20$ and $N=\frac{L}{2}-1$ (used in Fig.~\ref{fig:3-panel}), the asymptotic expansion yields $\braket{S_A}^0_{n=\frac{L-2}{2L}}\approx6.607$. Due to the slight deviation from $n=1/2$ (half-filling), the approximation to the exact value of $6.253$ is reasonable but not great. The latter can be evaluated as a sum, as explained in Eq.~(45) of~\cite{bianchi2021volume}, and will be used for the evaluation of $s_A$ instead. In the case of exactly half-filling, or far away from half-filling, one may use the asymptotic expansion as an excellent approximation.

In Fig.~\ref{fig:3-panel}(c) we show our entanglement probe $s_A$ from Eq.~\eqref{eq:smallS}. Again, we find a distinct region of highly entangled states that matches the outlines of the corresponding low-$\mathrm{IPR}_\mathrm{block}$ and GOE-level-spacing regions in Figs.~\ref{fig:3-panel}(a) and \ref{fig:3-panel}(b), including the nonthermal region close to the integrable line.


\section{Perturbation Theory and coarse grained block structure} \label{sec:Schrieffer}

In order to analyze the transition between exact fragmentation and large but finite interaction strength, we apply perturbation theory in the form of a Schrieffer-Wolff transformation \cite{bravyi_schrieffer-wolff_2011} to our model.
For a detailed derivation we refer to Appendix \ref{App:Schrieffer-Wolff}.

We find that exact degeneracy between blocks related by translation symmetry can lead to discontinuities in the exact eigenstates at $\epsilon= 1/V_1 =0$ (see Fig. \ref{fig:Schrieffer}). This is to be expected whenever the fragmented block structure does not respect symmetries of the model. The result is that product states of one block may dynamically evolve into a state that is partially or completely contained in a symmetry-related block with identical quantum numbers. Unlike the amplitude for scattering into other bond sectors, which is arbitrarily small, this type of transition can only be delayed to arbitrarily late times $\sim 1/\epsilon^2=V_1^2$. Experimentally, it can only be detected by resolving local particle number and prior knowledge of the block structure.

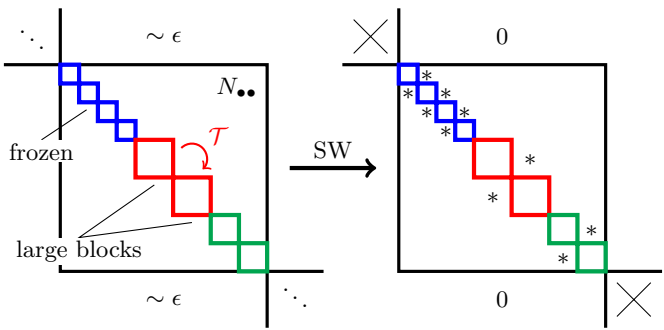
\begin{figure}[t]
    \centering
    \begin{tikzpicture}
    
    \begin{scope}
    \draw (1.625,.375) node{$\sim \epsilon$};
    \draw (1.625,-3.125) node{$\sim \epsilon$};
    \draw (2.6,-.3) node{$N_{\bullet\bullet}$};
    \draw (-.125,.435) node{$\ddots$};
    \draw (3.375,-3) node{$\ddots$};
    \draw[very thick] (-.5,0) -- (3,0) -- (3,-3.5);
    \draw[very thick] (.25,.75) -- (.25,-2.75) -- (3.75,-2.75);
    \draw[ultra thick,blue] (.25,0) rectangle (.5,-.25) rectangle (.75,-.5) rectangle (1,-.75) rectangle (1.25,-1);
    \draw[ultra thick,red] (1.25,-1) rectangle (1.75,-1.5) rectangle (2.25,-2);
    \draw[red,thick,->] (2,-1.25) node[xshift=.35cm,yshift=.3cm]{$\mathcal{T}$} +(135:.2) arc(135:-45:.2);
    \draw[ultra thick,Green] (2.25,-2) rectangle (2.625,-2.375) rectangle (3,-2.75);
    \draw  (.6,-.6) -- (0,-1) node[below,fill=white,inner sep=1pt]{frozen};
    \draw (1.5,-1.6) -- (.5,-2.3) node[below,fill=white,inner sep=1pt]{large blocks} -- (2,-2.1);
    \end{scope}

    \draw[ultra thick,->] (3.3,-1.375) -- node[above]{SW} (4.45,-1.375);

    \begin{scope}[xshift=4.5cm]
    \draw (.1,.15) -- (-.35,.6) (.1,.6) -- (-.35,.15);
    \draw (1.625,.375) node{$0$};
    \draw (1.625,-3.125) node{$0$};
    \begin{scope}[xshift=3.5cm,yshift=-3.5cm]
        \draw (.1,.15) -- (-.35,.6) (.1,.6) -- (-.35,.15);
    \end{scope}
    \draw[very thick] (-.5,0) -- (3,0) -- (3,-3.5);
    \draw[very thick] (.25,.75) -- (.25,-2.75) -- (3.75,-2.75);
    \draw[ultra thick,blue] (.25,0) rectangle (.5,-.25) rectangle (.75,-.5) rectangle (1,-.75) rectangle (1.25,-1);
    \draw[ultra thick,red] (1.25,-1) rectangle (1.75,-1.5) rectangle (2.25,-2);
    \draw[ultra thick,Green] (2.25,-2) rectangle (2.625,-2.375) rectangle (3,-2.75);
    \draw (.625,-.125) node[yshift=-.6mm]{*} (.875,-.375) node[yshift=-.6mm]{*} (1.125,-.625) node[yshift=-.6mm]{*};
    \draw (2,-1.25) node[yshift=-.6mm]{*} (1.5,-1.75) node[yshift=-.6mm]{*};
    \fill (2.8125,-2.1875) node[yshift=-.6mm]{*} (2.4375,-2.5625) node[yshift=-.6mm]{*};
    \begin{scope}[xshift=-.25cm,yshift=-.25cm]
    \draw (.625,-.125) node[yshift=-.6mm]{*} (.875,-.375) node[yshift=-.6mm]{*} (1.125,-.625)  node[yshift=-.6mm]{*};
    \end{scope}
    \end{scope}
    
    \end{tikzpicture}
    \caption{We illustrate a finite perturbation to the ideal block structure before and after applying a Schrieffer-Wolff (SW) transformation. Displayed is a single bond sector defined by the bond number $N_{\bullet \bullet}= \sum_{x}\hat{n}_x\hat{n}_{x+1}$ with its internal block structure of frozen states and large blocks, marked in colored squares. Due to $N_{\bullet \bullet}$ being only approximately conserved, there are block-off-diagonal terms of order $\epsilon=1/V_1$. $\mathcal{T}$ represents the  action of the translation operator, where blocks that are related by translation share the same color. After applying SW, different bond sectors become disconnected and acquire additional matrix elements as indicated by $*$ symbols, which predict discontinuities in the exact eigenstates due to higher-order perturbative effects.}
    \label{fig:Schrieffer}
\end{figure}

\begin{figure*}[t!]
    \centering
    \includegraphics[width=.8\textwidth]{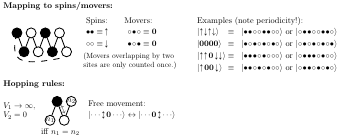}
    \caption{Spin-mover mapping \cite{Dias_2000} and effective hopping rules \cite{frey_hilbert_2022} for the fragmented regimes of the extended Fermi-Hubbard model. In the limit of $V_1 \to \infty$ and $V_2=0$ large blocks are comprised of states that host one ore more movers that hop by changing position with a neighboring spin. When all movers traverse the entire system once (periodic boundary conditions assumed) the resulting state corresponds to a spatial translation by two sites in the fermionic Fock state.}
    \label{fig:spin-mover}
\end{figure*}

In the following we show that translation symmetry can be used in order to determine whether discontinuous mixing of eigenstates occurs at $\epsilon=0$ for the case $V_1 \to \infty$. We make use of the spin-mover mapping that has been used in the past to study this model \cite{Dias_2000,de2019dynamics,frey_hilbert_2022} (see Fig.~\ref{fig:spin-mover}). Consider a block $A$ that is characterized by a nonzero number of movers, \ie, not a frozen state. We can conclude based on the spin-mover dynamics that for any product state $\ket{a}$ in said block the state $\hat{T}^2 \ket{a}$ is also contained in the same block. Therefore we can group the product states in $A$ into closed orbits under $\hat{T}^2$. Consider one such set $\ket{a,1},\ket{a,2},\dots$. Applying $\hat{T}$ to each state yields a corresponding set  $\ket{b,1},\ket{b,2},\dots$ in the block $B$ whose basis states are the translated states of $A$. Note that $\hat{T}\ket{a,n}=\ket{b,n}$ and $\hat{T}\ket{b,n}=\ket{a,n+1}$. We can construct translation eigenstates by first constructing $\hat{T}^2$ eigenstates with eigenvalue $t^2$, $\ket{\psi_{A/B},t^2} = \sum_{n=1}^N (\Bar{t}^2)^{n-1} \ket{a/b,n}$ and then taking the linear combination $\ket{\psi_{A},t^2} + \Bar{t} \ket{\psi_{B},t^2} $. Using $\hat{T} \ket{\psi_{A},t^2} = \ket{\psi_{B},t^2}$ and $\hat{T} \ket{\psi_{B},t^2} = t^2 \ket{\psi_{A},t^2}$, one can easily verify that these are translation eigenstates with $\hat{T}$ eigenvalue $t$. In the fragmented limit $\epsilon=0$, the eigenstates of $\hat{H}$ can be chosen to be linear combinations of product states purely in $A$ or $B$. Since $A$ and $B$ are invariant under $\hat{T}^2$, we can choose simultaneous eigenstates $\widetilde{\ket{\psi_{A},t^2}}$ of $\hat{H}$ and $\hat{T}^2$ that generally are combinations of several $\hat{T}^2$ eigenstates $\{\ket{\psi_{A},t^2}_j\}$ constructed in the fashion outlined above. Applying $\hat{T}$ yields a degenerate state $\widetilde{\ket{\psi_{B},t^2}}$ in block $B$ and simultaneous diagonalization of $\hat{H}$ and $\hat{T}$ requires equal weight superpositions of these states $\ket{\psi,t} \propto \widetilde{\ket{\psi_{A},t^2}} + \Bar{t} \widetilde{\ket{\psi_{B},t^2}}$. We recover the block localized eigenstates via $\widetilde{\ket{\psi_{A/B},t^2}} \propto \ket{\psi,t} \pm \ket{\psi,-t} $. Suppose there is no discontinuous mixing of eigenstates, i.e., there exists a well-defined perturbative expansion $\widetilde{\ket{\psi_{A},t^2}}_{\epsilon>0} \propto \widetilde{\ket{\psi_{A},t^2}} + \epsilon \cdots $. Applying $\hat{T}$ yields the corresponding degenerate perturbative eigenstate $\widetilde{\ket{\psi_{B},t^2}}_{\epsilon>0}$ that is mostly localized on $B$ and the hence by the same construction the combined $\hat{H}$ and $\hat{T}$ eigenstates $\widetilde{\ket{\psi,\pm t}}$ must be degenerate. We also conclude that a perturbative expansion for all eigenstates exists if and only if all simultaneous $\hat{H}$ and $\hat{T}$ eigenstates can be sorted into exactly degenerate pairs with $\hat{T}$ eigenvalues $\pm t$.

Next we consider systems where $L$ is a multiple of 4. Here we expect large blocks to contain states with  $t^2=-1$, which can be combined with translated eigenstates to give $\hat{T}$ eigenvalues $\pm i$.  By construction $[\hat{H}_\mathrm{eff},\hat{T}]=0$, $[\hat{H}_\mathrm{eff},\hat{P}]=0$ and $\hat{P} \hat{T} \hat{P} = \hat{T}^\dagger$, which implies degeneracy between $i$ and $-i$ translation eigenvalues. Therefore $\bra{\psi_{B},t^2}\hat{H}_\mathrm{eff} \ket{\psi_{A},t^2} = 0$ to all orders in $\epsilon$ and there is no discontinuous mixing of these states. We find numerically that these are the only translation eigenstates that can be perfectly matched into degenerate pairs and the block localization of eigenstates can be preserved at finite $\epsilon>0$.

Additionally, there exist an extensive number of frozen states in the fragmented limit, which are degenerate even if not related by symmetry. In order to avoid numerical discontinuities at, say, $V_1 \to \infty$, we replace the naive block structure obtained from $\hat{H}_\infty^{(1)}$ with a coarse grained one. The latter is obtained by combining blocks that are either related by translation symmetry or are frozen states within the same bond sector.


\section{Application: Wannier-Stark problem}

So far we have demonstrated the block IPR for one particular model.
In order to provide evidence for the claim that our approach is universal and indeed applies to any family of Hamiltonians that have a fragmented limit, we now apply the same methods to another, popular type of fragmented model. 

We consider a spinless fermionic chain with open boundary conditions and constant transverse electric field, also known as the Wannier-Stark problem
\cite{RevModPhys.34.645},

\begin{equation}\label{eq:dipoleHam}
\begin{aligned}
    \hat{H}=&-t\sum_x (\hat{c}^\dagger_{x+1}\hat{c}_x+\mathrm{H.c.}) \\ 
    &+ V_1\sum_{x}\hat{n}_x\hat{n}_{x+1} + E \sum_x x \, \hat{n}_x \,.
\end{aligned}
\end{equation}

Again setting $t=1$, we now take $V_1 \sim 1$ and $E \gg 1$. Since the tight binding term has no block-diagonal component with respect to eigenspaces of the dipole moment $\hat{D} = \sum_x x \, \hat{n}_x$, one needs to go to second order in perturbation theory to arrive at an effective Hamiltonian that is proportional to $V_1$ and whose nondiagonal component is described by pair-hopping processes \cite{Moudgalya_2021},

\begin{equation}\label{eq:effective_dipoleHam}
\begin{aligned}
    \frac{\hat{H}_\mathrm{eff}}{V_1} = &-\frac{1}{E^2} \sum_x (\hat{c}^\dagger_{x} \hat{c}^\dagger_{x+3} \hat{c}_{x+2} \hat{c}_{x+1} + \mathrm{h.c.}) \\ 
    &+ \left( 1 - \frac{2}{E^2} \right) \sum_{x}\hat{n}_x\hat{n}_{x+1} + \frac{2}{E^2} \sum_{x}\hat{n}_x\hat{n}_{x+2} 
\end{aligned}
\end{equation}

From this it is clear that in the limit $E \to \infty$ the bond number $\hat{N}_{\bullet \bullet} = \sum_{x}\hat{n}_x\hat{n}_{x+1}$ is conserved along with the dipole moment $\hat{D}$. The finite perturbation in the vicinity of the fragmented limit, however, is now given by the dipole conserving pair-hopping. These types of models have been studied in the past and are known to be fragmented \cite{PhysRevX.10.011047,PhysRevB.101.214205}. We can apply the same arguments and techniques outlined in Secs.~\ref{sec:extendedFermi} and \ref{sec:Schrieffer} in order to define an effective block structure with respect to which we compute the block IPR as a function of $E$. Since nearest-neighbor interactions are crucial in order to obtain dipole conserving hopping in this manner, we find that the block structure is somewhat different from the one obtained in other studies which only consider a pair-hopping term. It is defined by the connected components of the block-diagonal part of pair hopping with respect to eigenspaces of $\hat{N}_{\bullet \bullet}$. In other words, the relevant adjacency matrix is given by

\begin{equation}\label{eq:dipoleblockHam}
\begin{aligned}
    \hat{H}_\mathrm{block} = \sum_x \hat{P}_x \, (\hat{c}^\dagger_{x} \hat{c}^\dagger_{x+3} \hat{c}_{x+2} \hat{c}_{x+1} + \mathrm{H.c.}) \, \hat{P}_x \, ,
\end{aligned}
\end{equation}
where $\hat{P_x} = (\hat{n}_{x-1} - \hat{n}_{x+4})^2 $ is a projection operator. Here we define $\hat{n}_0 = \hat{n}_\mathrm{L+1} =0$.

In Fig.~\ref{fig:dipole-panel}(a) we show the level statistics, entanglement entropy, block IPR and variance of $\mathrm{IPR}_\mathrm{block}$ as a function of $E$ for fixed $V_1=1$. Due to lack of translational symmetry, we are limited to $L=16$ and half-filling. Similarly to the extended Fermi-Hubbard model, all three probes indicate the existence of two distinct regimes. We find that the onset of fragmentation seems to coincide with the distinct drop in the level ratio and entanglement entropy. Figure~\ref{fig:dipole-panel}(b) shows the corresponding scaling analysis for the midpoint of $\mathrm{IPR}_\mathrm{block}$ and peak variance of $\mathrm{IPR}_\mathrm{block}$ respectively. We find overall less agreement between the two predicted values of $E^*$, likely due to the limitation in system size.

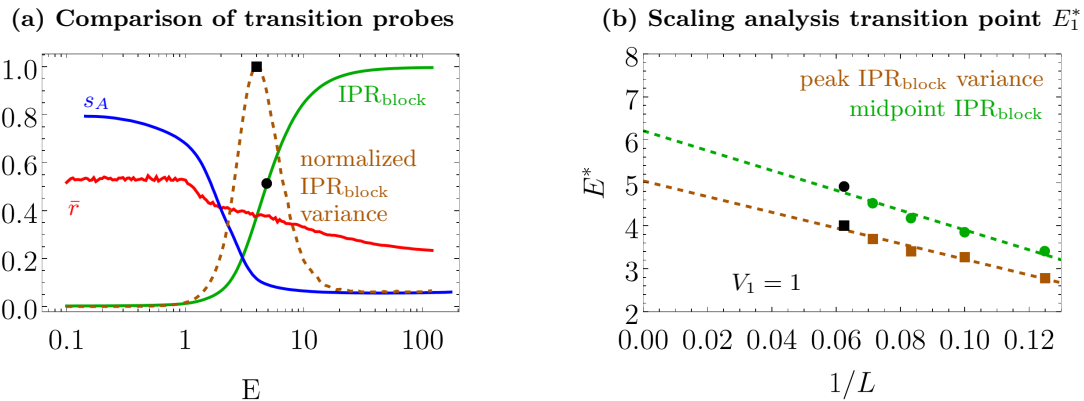
\begin{figure*}[t!]
    \centering
    \begin{tikzpicture}
        \begin{scope}[yshift=-5mm]
        \draw (-4,-5) node{\includegraphics[width=7cm]{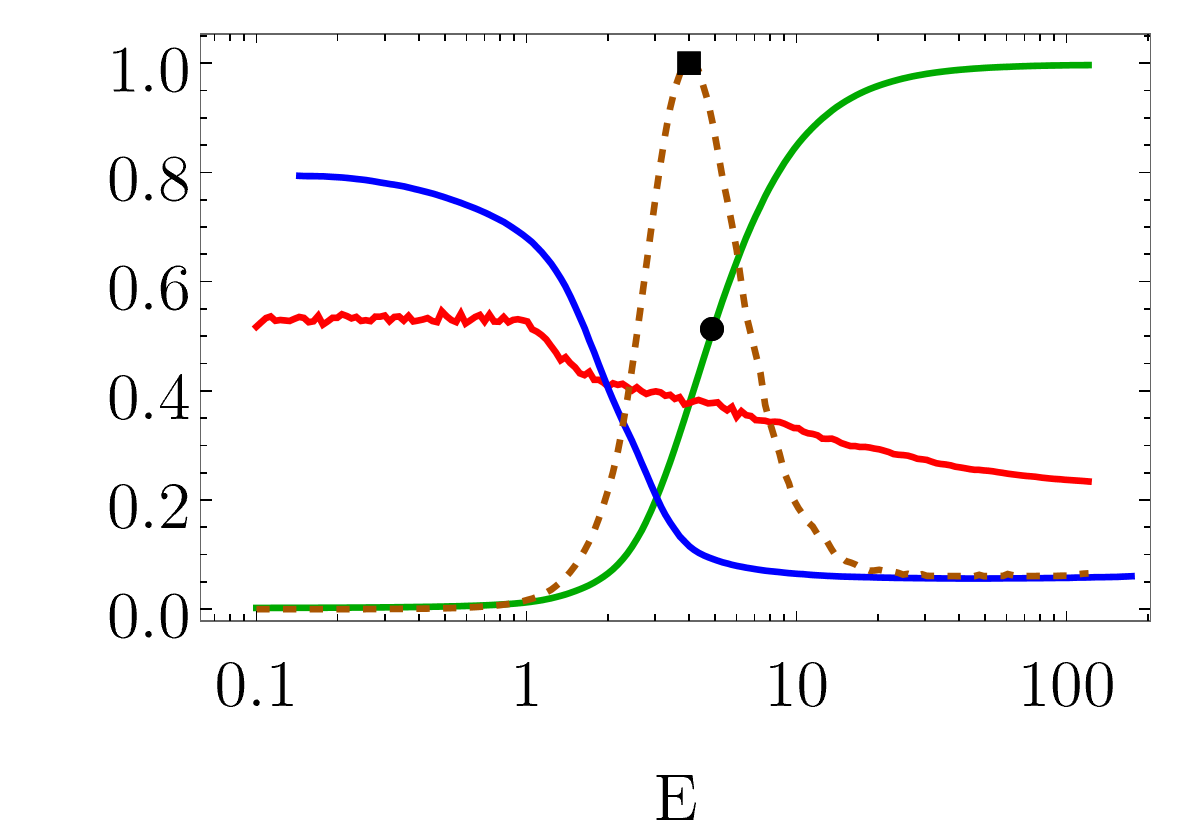}};
        \draw (-3.8,-2.3) node{\textbf{(a) Comparison of transition probes}};
        \draw (-6.1,-4.8) node[red,right]{$\bar{r}$} (-1.8,-3.3) node[dgreen]{$\mathrm{IPR}_\mathrm{block}$} (-5.9,-3.4) node[blue,right]{$s_A$} (-1.8,-4.5) node[dorange,text width=2.2cm]{normalized $\mathrm{IPR}_\mathrm{block}$ variance};
        \draw (4,-5) node{\includegraphics[width=7cm]{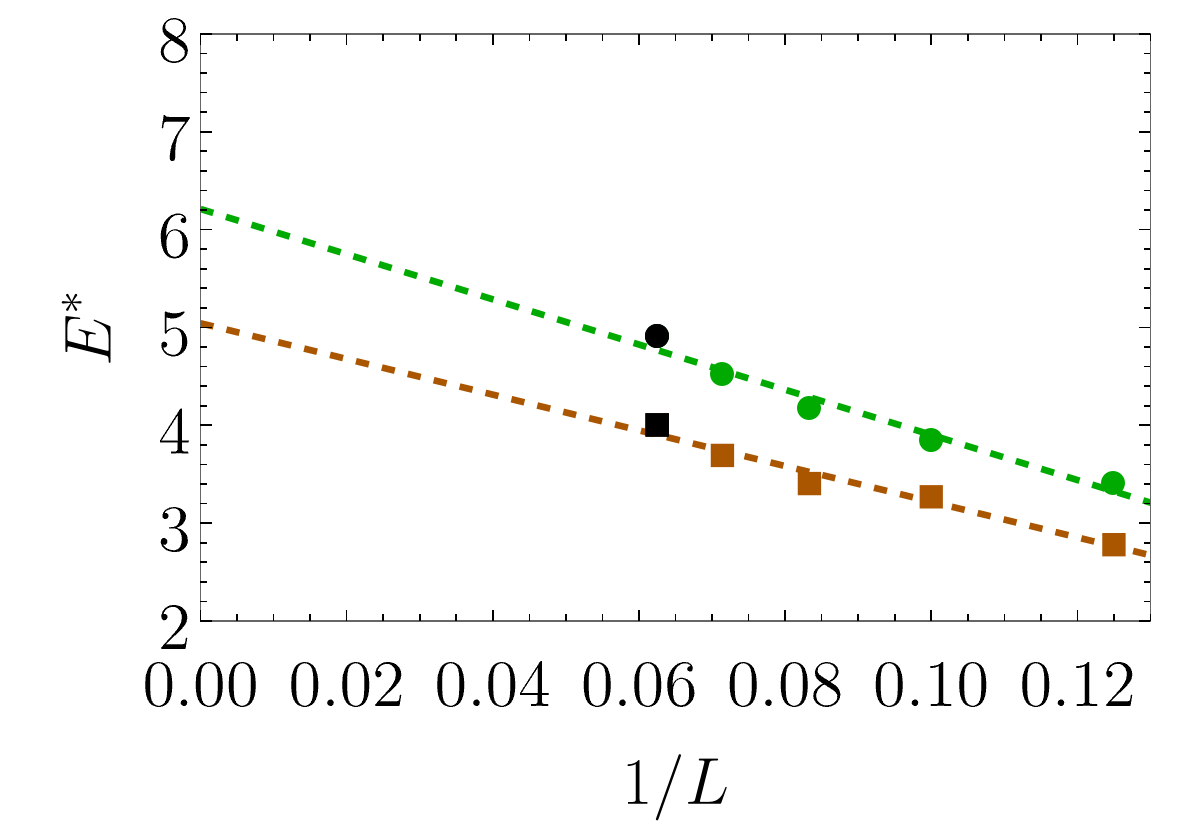}};
        \draw (4.3,-2.3) node{\textbf{(b) Scaling analysis transition point $E^*_1$}};
        \draw (3.3,-5.8) node{$V_1=1$} (4.4,-4);
        \draw (7.1,-3.5) node[left,dgreen]{midpoint $\mathrm{IPR}_\mathrm{block}$} (7.1,-3.1) node[left,dorange]{peak $\mathrm{IPR}_\mathrm{block}$ variance};
        \end{scope}
    \end{tikzpicture}
\caption{\textbf{(a)} Comparison of $\mathrm{IPR}_\mathrm{block}$, level statistics, and entanglement entropy for the $L=16$ Wannier-Stark Hamiltonian [Eq.~(\ref{eq:dipoleHam})] at half-filling and $V_1=1$. The black circle indicates the midpoint of the jump in average $\mathrm{IPR}_\mathrm{block}$, while the black square indicates the position of maximal variance in $\mathrm{IPR}_\mathrm{block}$ across eigenstates.  \textbf{(b)} Scaling analysis of the respective values for $E^*$ obtained in \textbf{(a)} across varying system size and for $V_1=1$. }
    \label{fig:dipole-panel}
\end{figure*}


\section{Discussion and outlook}

We studied the possibility that fragmentation may be an identifiable phase
with a suitable order parameter. As we will summarize in the following, we have introduced a numerical probe, the \emph{block inverse participation ratio} $\mathrm{IPR}_\mathrm{block}$, as such an order parameter, allowing us to
quantify the degree to which a parameter-dependent model is fragmented. We do not argue in terms of perturbative effects on the lifetime of fragmenting constraints. Instead, $\mathrm{IPR}_\mathrm{block}$ and our other probes (level-spacing, entanglement entropy) directly test the properties of the exact eigenstates for a continuum of model parameters.

Using perturbation theory, we were able to deal with degeneracy to ensure that the $\mathrm{IPR}_\mathrm{block}$ is well-defined when using numerical exact diagonalization. Comparing this $\mathrm{IPR}_\mathrm{block}$ with more conventional probes of ergodicity like level-spacing statistics and entanglement entropy, we were able to show that its prediction of a transition between nonfragmented and fragmented regimes is exactly in line with the predicted transition between thermal and non-thermal regimes at strong interactions based on level statistics and entanglement entropy.

Additional nonthermal regimes in proximity to the integrable limit of the extended Fermi-Hubbard model are not indicated by $\mathrm{IPR}_\mathrm{block}$. This shows that it is a numerical probe that specifically detects and quantifies fragmentation as opposed to general mechanisms of ergodicity breaking. It therefore explains the mechanism behind the transitions/crossovers in those parts of the phase diagram where its predictions agree with the generic probes.

Based on finite-size scaling, we were also able to show that it is robust in the thermodynamic limit and defines a regime of proximity to the exactly fragmented limit that is dominated by the effects of (approximate) fragmentation. We demonstrate that $\mathrm{IPR}_\mathrm{block}$ can be used in rather general settings that shows a fragmented limit by applying the same approach to a dipole conserving model. Let us emphasize that defining and calculating the block structure for such models is numerically very inexpensive, as it only requires finding connected components of a sparse matrix, for which there are efficient algorithms. In practice, one may first look at the conventional IPR with respect to a local product basis, in the case of a finite system, or generally any universal probe such as level statistics or entanglement entropy in order to identify potentially fragmented, non-thermal regimes. Once such appropriate limits are identified, one finds the (approximate) adjacency matrix in order to compute $\mathrm{IPR}_\mathrm{block}$ and thereby identify the cause of ergodicity breaking as either fragmentation or through some other mechanism.

While infinite interaction strengths are not experimentally realizable, very large but finite ones most certainly are accessible in quantum simulators or quantum computers. The effect and methods presented in this paper can therefore be studied in experiments. Our results suggest that fragmentation is not restricted to a low-dimensional subset (hypersurface) of Hamiltonians in the thermodynamic limit, but rather represents a finite volume around the exactly fragmented regime and may constitute a separate phase with $\mathrm{IPR}_\mathrm{block}$ serving as the order parameter.

\section*{Acknowledgments}
L.H. acknowledges support by the Alexander von Humboldt Foundation. S.R. acknowledges support from the Australian Research Council through Grant No. DP200101118. This research was supported by The University of Melbourne’s Research Computing Services and the Petascale Campus Initiative.


\newpage
\bibliography{references}

\clearpage
\onecolumngrid
\appendix

\section{Schrieffer-Wolff transformation} \label{App:Schrieffer-Wolff}


Without loss of generality we consider the case where $V_1\gg 1$.

We begin by re-scaling $\hat{H} \rightarrow \hat{H}/V_1$ and splitting the Hamiltonian into $\hat{H}_0(\epsilon) + \epsilon^* \hat{H}_\mathrm{pert}$ with $\epsilon = 1/V_1$,

\begin{align}
    \hat{H}_0 &= \hat{N}_{\bullet\bullet} + \epsilon  \hat{H}^{(1)}_\infty \\
    \hat{H}_\mathrm{pert} =& \left(- \sum_x (\hat{c}^\dagger_{x+1}\hat{c}_x+\mathrm{h.c.}) -  \hat{H}^{(1)}_\infty \right) .
\end{align}

\noindent The unperturbed eigenstates of $\hat{H}_0$ can be indexed by the eigenvalue of $\hat{N}_{\bullet\bullet}$ and an additional index $i$,

\begin{align}
    \hat{H}_0 \ket{i, N_{\bullet\bullet}} = (N_{\bullet\bullet}+E^{N_{\bullet\bullet}}_i(\epsilon)) \ket{i, N_{\bullet\bullet}} .
\end{align}

\noindent We can write $E^{N_{\bullet\bullet}}_i(\epsilon) =e^{N_{\bullet\bullet}}_i \cdot \epsilon$.
These eigenstates are independent of $\epsilon > 0$ and can be made continuous at $\epsilon=0$, since each block takes the form $N_{\bullet\bullet} \cdot \hat{\mathbb{1}} + \epsilon \cdot \hat{X}$, with $\hat{X}$ independent of $\epsilon$.

The effective Hamiltonian projected onto a manifold $N_{\bullet\bullet}$ up to corrections of order $\epsilon^4$ may be written as

\begin{equation}
\begin{aligned}
   \bra{i} \hat{H}_\mathrm{eff}^{N_{\bullet\bullet}} \ket{j} =  &(N_{\bullet\bullet}+E^{N_{\bullet\bullet}}_i(\epsilon)) \delta_{i,j} + \epsilon \cdot 0 + \\ \epsilon^2 \sum_k \left\{\vphantom{\left[ 1 - \epsilon \left(\frac{e^{N_{\bullet\bullet}}_i + e^{N_{\bullet\bullet}}_j}{2} - e^{N_{\bullet\bullet}-1}_k \right) \right] } \right. &\bra{i,N_{\bullet\bullet}} \hat{H}_\mathrm{pert} \ket{k,N_{\bullet\bullet}-1} \bra{k,N_{\bullet\bullet}-1} \hat{H}_\mathrm{pert} \ket{j,N_{\bullet\bullet}} \left[ 1 - \epsilon \left(\frac{e^{N_{\bullet\bullet}}_i + e^{N_{\bullet\bullet}}_j}{2} - e^{N_{\bullet\bullet}-1}_k \right) + \mathcal{O}(\epsilon^2) \right] 
   - \\ &\left. \bra{i,N_{\bullet\bullet}} \hat{H}_\mathrm{pert} \ket{k,N_{\bullet\bullet}+1} \bra{k,N_{\bullet\bullet}+1} \hat{H}_\mathrm{pert} \ket{j,N_{\bullet\bullet}} \left[ 1 + \epsilon \left(\frac{e^{N_{\bullet\bullet}}_i + e^{N_{\bullet\bullet}}_j}{2} - e^{N_{\bullet\bullet}+1}_k \right) + \mathcal{O}(\epsilon^2) \right] \right\} \\
   &+ 0 \cdot \epsilon^3    .
\end{aligned}   
\end{equation}

\noindent Using the relations

\begin{equation}
H_\mathrm{eff}= \hat{T} \hat{H} \hat{T}^\dagger\,,\qquad
\hat{T} = e^{i\hat{S}}\,,\qquad
\hat{S} = \epsilon S_1 + \epsilon^2 \hat{S}_2 + \dots \, ,
\end{equation}
we find that the eigenvectors of $\hat{H}$ can be computed by first finding the eigenvectors of $\hat{H}_\mathrm{eff}$ and then applying $\hat{T}^\dagger = \hat{\mathbb{1}} - i \epsilon \hat{S}_1 - i \epsilon^2 \hat{S}_2 - \epsilon^2 \hat{S}_1^2 + \dots $ . The eigenvectors of $\hat{H}_\mathrm{eff}$ can in turn be expanded perturbatively unless degeneracies are only lifted by the diagonal corrections, i.e., at order $\epsilon^2$, while off-diagonal terms couple the degenerate states. In particular, there are exactly degenerate diagonal entries to all orders due to translation symmetry and the noninvariance of the fragmented block structure under translation. Note that $\hat{S}$ is chosen to be purely off-diagonal with respect to the different bond sectors.
The matrix elements of $\hat{S}_1$ are given by

\begin{equation}
\begin{aligned}
& \left\langle i, N_{\bullet\bullet} \pm 1  \left|i \epsilon \hat{S}_1\right| j, N_{\bullet\bullet} \right\rangle=\frac{\langle i, N_{\bullet\bullet} \pm 1 |\epsilon  \hat{H}_\mathrm{pert}| j, N_{\bullet\bullet} \rangle}{E^{N_{\bullet\bullet}\pm1}_i(\epsilon)-E^{N_{\bullet\bullet}}_i(\epsilon)} ,    \\
& \left\langle i, N_{\bullet\bullet} \left|i \epsilon \hat{S}_1\right| j, N_{\bullet\bullet}\right\rangle=0 .
\end{aligned}
\end{equation}

\noindent A naive perturbative expansion of the eigenstates takes the form

    
\begin{equation}
\begin{aligned}
    \ket{i, N_{\bullet\bullet}}_\epsilon = &\ket{i, N_{\bullet\bullet}}_0 \\
    &- i \epsilon \hat{S}_1 \ket{i, N_{\bullet\bullet}}_0 - i \epsilon^2 \hat{S}_2 \ket{i, N_{\bullet\bullet}}_0 \\
    & + \epsilon^2 \sum_{j \neq i} \frac{\bra{j, N_{\bullet\bullet}}_0 \hat{H}_\mathrm{eff,2}^{N_{\bullet\bullet}}\ket{i, N_{\bullet\bullet}}_0}{E^{N_{\bullet\bullet}}_i(\epsilon)-E^{N_{\bullet\bullet}}_j(\epsilon)} \ket{j, N_{\bullet\bullet}}_0 - \epsilon^2 \hat{S}_1^2 \ket{i, N_{\bullet\bullet}}_0 + \dots
\end{aligned}
\end{equation}

\noindent The corrections appearing in line two only mix in states from different bond sectors, while those in line three can mix in states from other blocks within the same bond sector. Notably, $E^{N_{\bullet\bullet}}_i(\epsilon)-E^{N_{\bullet\bullet}}_j(\epsilon) = \epsilon (e^{N_{\bullet\bullet}}_i-e^{N_{\bullet\bullet}}_j)$ and so the first term in row three is a first-order correction in $\epsilon$ unless the matrix element of $\hat{H}_\mathrm{eff,2}^{N_{\bullet\bullet}}$ between degenerate eigenstates, in particular in different blocks related by $\hat{T}$ symmetry, is finite. In this case we find a discontinuity in the eigenstates at $\epsilon=0$. This results in a transition between blocks on a timescale $\sim 1/\epsilon^2=V_1^2$.

\noindent These types of discontinuities are generally expected when the Hamiltonian takes the following form: 

\begin{enumerate}
    \item $\hat{H} = \epsilon^{-1} \hat{D} + \hat{h}$, with $\hat{D}$ diagonal in product basis
    \item $\hat{h} = \hat{h}^\mathrm{d} + \hat{h}^\mathrm{nd}$ with $[\hat{h}^\mathrm{d},\hat{D}]=0$ and $[\hat{h}^\mathrm{nd},\hat{D}]=0$
    \item $\epsilon \hat{H} = \hat{D} + \underbrace{\epsilon \hat{h}^\mathrm{d}}_\mathrm{strong~pert} + \underbrace{\epsilon \hat{h}^\mathrm{nd}}_\mathrm{pert}$ 
    \item $\hat{H_0} = \hat{D} + \epsilon \hat{h}^\mathrm{d}$ is fragmented, i.e., in general $\bra{d,1} \hat{h}^\mathrm{d} \ket{d,2}=0$ for $\ket{d,1}, \ket{d,2}$ in same eigenspace of $\hat{D}$
    \item A symmetry $\hat{S}$ of $\hat{H}$ does not leave the block structure invariant: $\bra{d,1} \hat{S} \ket{d,2} \neq 0$
    \item The effective Hamiltonian couples degenerate states at some order, i.e., $\bra{d,1} \hat{H}_\mathrm{eff}^{(n)} \ket{d,2} \neq 0$. It is sufficient if $\bra{d,3} \hat{H}_\mathrm{eff}^{(n)} \ket{d,1} \neq 0$, where $\bra{d,3} (\hat{h}^\mathrm{d})^k \ket{d,2} \neq 0$ for some $k \in \mathbb{N}$.
\end{enumerate}


\bigskip
\section{Additional data}\label{App:MorePlots}

In Fig.~\ref{fig:additional-data} we present the individual plots for $\mathrm{IPR}_\mathrm{block}$ with respect to $V_1$ and with respect to $V_2$ across the entire range of $V_1$ and $V_2$ values. It is evident that there is some amount of overlap between the respective limiting block structures. This is indicated by the slight increase in $V_1$-$\mathrm{IPR}_\mathrm{block}$ when approaching the $V_2 \to \infty$ limit and vice versa. Due to this overlap one may identify the thermal region by only considering one of these IPRs.
We also show data for the level-spacing statistics at half-filling, here for $L=18$, $N=9$. As outlined in Sec.~\ref{sec:LevelSpacing}, one expects generalized GOE statistics in this case because of the presence of particle-hole symmetry. We find this to be the case and also note the washing out of the outline and features of the thermal to nonthermal transition when compared to the level statistics away from half-filling (compare Fig.~\ref{fig:3-panel}). We observe much less difference between different filling fractions in the entanglement entropy and $\mathrm{IPR}_\mathrm{block}$.

\bigbreak
\begin{figure*}[h]
    \centering
    \begin{tikzpicture}
        \draw (-6.2,2.6) node{\textbf{(a) Block IPR w.r.t. $V_1$}};
        \draw (-.2,2.6) node{\textbf{(b) Block IPR w.r.t. $V_2$}};
        \draw (5.7,2.6) node{\textbf{(c) Level-spacing at half-filling}};
        \draw (-6,0) node{\includegraphics[height=4.6cm]{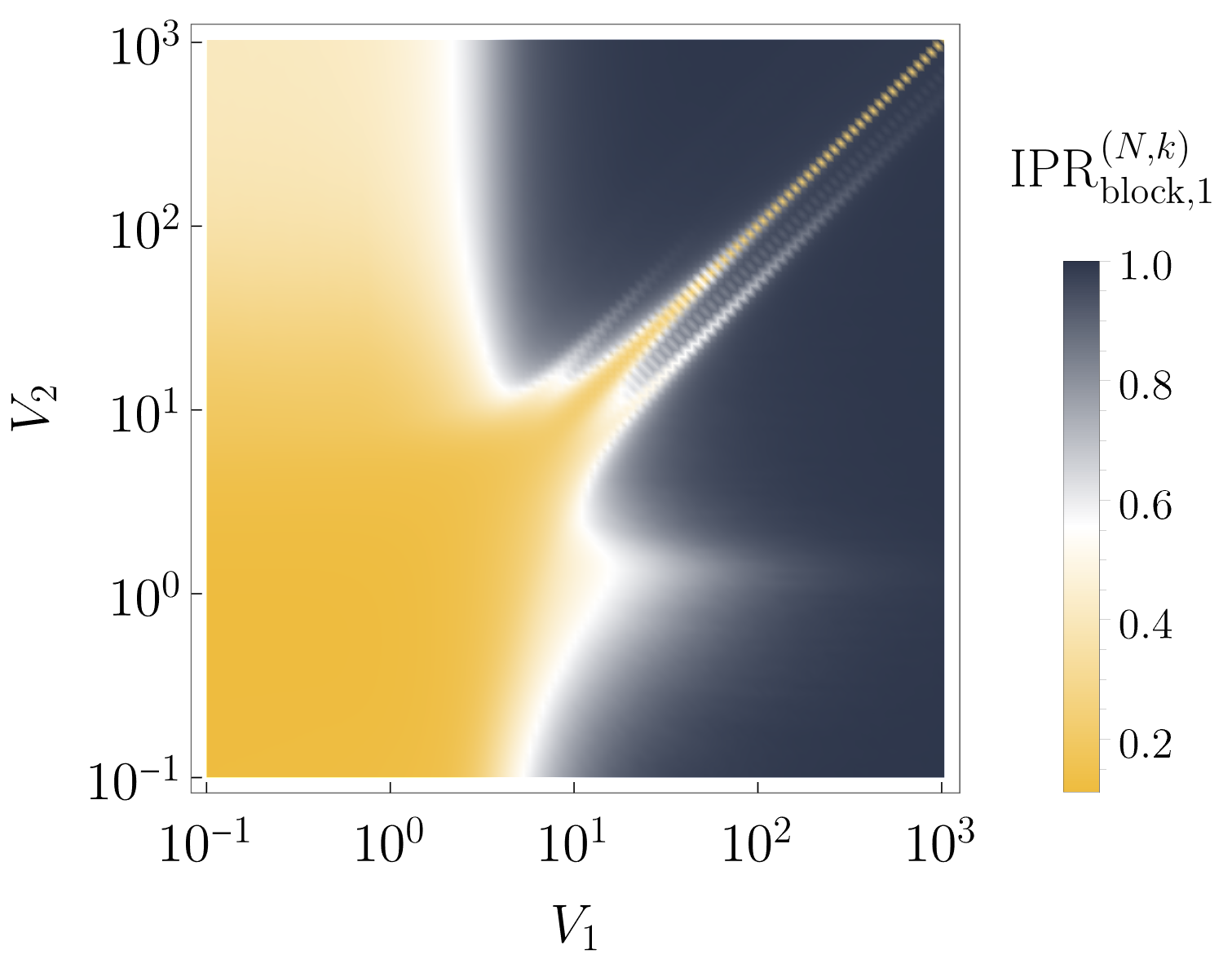}};
        \draw (0,0) node{\includegraphics[height=4.6cm]{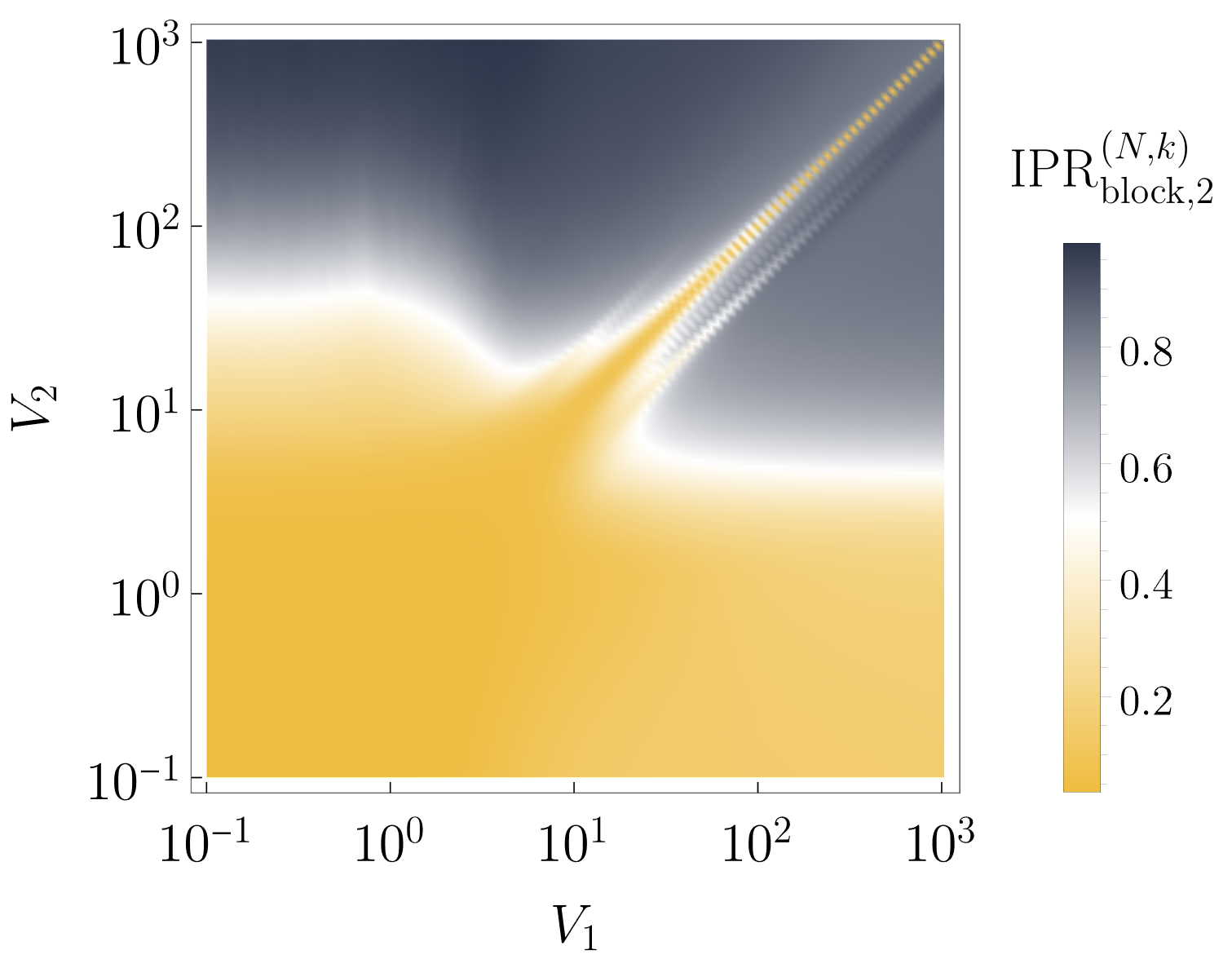}};
        \draw (6,0) node{\includegraphics[height=4.6cm]{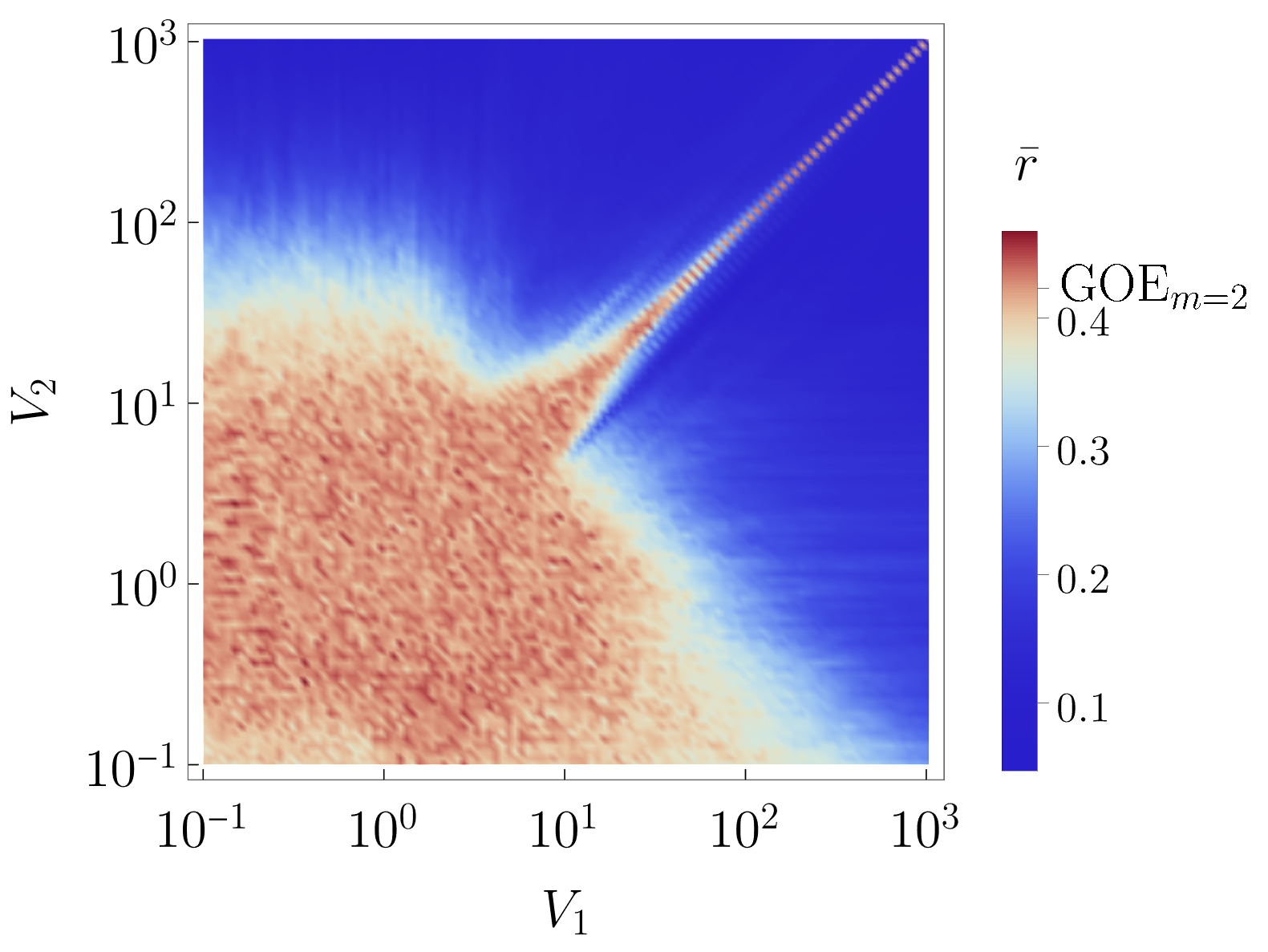}};
        \node [fill=white,rounded corners=2pt,inner sep=1pt] at (-6.2,1.8) {$L=20,N=9,k=1$};
        \node [fill=white,rounded corners=2pt,inner sep=1pt] at (-.2,1.8) {$L=20,N=9,k=1$};
        \node [fill=white,rounded corners=2pt,inner sep=1pt] at (5.7,1.8) {$L=18,N=9,k=1$};
    \end{tikzpicture}
\caption{Individual plots for \textbf{(a)} block IPR with respect to $V_1$ and \textbf{(b)} block IPR with respect to $V_2$ across all values of $V_1,V_2$ [Figure \ref{fig:3-panel}\textbf{(a)} combines these plots] and \textbf{c)} level-spacing statistics for half-filling follows $\mathrm{GOE}_{m=2}$ within the thermal regime. We indicate system size $L$, particle number $N$ and momentum sector $k$.}
\label{fig:additional-data}
\end{figure*}
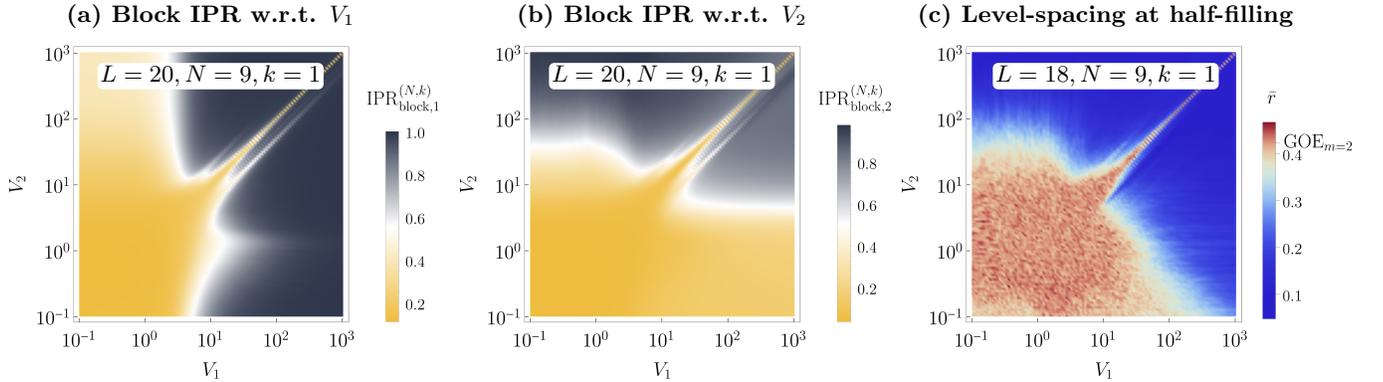


\end{document}